\pdfoutput=1
\documentclass[10pt,twocolumn,letterpaper]{article}

\usepackage[pagenumbers]{cvpr} %

\usepackage{times}
\usepackage{epsfig}
\usepackage{graphicx}
\usepackage{amsmath}
\usepackage{amssymb}
\usepackage{mathtools}

\usepackage{booktabs}
\usepackage{multirow}
\usepackage{caption}
\usepackage{xspace}
\usepackage[dvipsnames]{xcolor} %
\usepackage{siunitx}
\usepackage{algorithm}
\usepackage{algpseudocode}
\usepackage{multirow}

\usepackage{inconsolata}

\definecolor{cvprblue}{rgb}{0.21,0.49,0.74}
\usepackage[pagebackref,breaklinks,colorlinks,citecolor=cvprblue]{hyperref}

\usepackage[capitalize]{cleveref}
\crefname{section}{Sec.}{Secs.}
\Crefname{section}{Section}{Sections}
\Crefname{table}{Table}{Tables}
\crefname{table}{Tab.}{Tabs.}

\usepackage{algorithm}      %
\usepackage{algpseudocode}  %

\usepackage{fvextra}
\usepackage[frozencache,cachedir=minted-cache]{minted}
\newcommand{\ourtitle}{HSM: Hierarchical Scene Motifs for Multi-Scale Indoor Scene Generation}
\newcommand{\ours}{HSM\xspace}
\newcommand{\oursshort}{HSM\xspace}
\newcommand{\ourspossessive}{HSM's\xspace}
\newcommand{\oursfull}{Hierarchical Scene Motifs\xspace}

\newcommand{\inputdesc}{\ensuremath{T}\xspace}

\newcommand{\occthresh}{\ensuremath{t_{\text{occ}}}\xspace}
\newcommand{\iterthresh}{\ensuremath{t_{\text{iter}}}\xspace}
\newcommand{\clearancethresh}{\ensuremath{t_{\text{clear}}}\xspace}
\newcommand{\topheight}{\ensuremath{h_{\text{top}}}\xspace}

\newcommand{\normalthresh}{\ensuremath{t_{\text{norm}}}\xspace}
\newcommand{\adjacentthresh}{\ensuremath{t_{\text{adj}}}\xspace}
\newcommand{\wallprojthresh}{\ensuremath{t_{\text{wall}}}\xspace}
\newcommand{\segthresh}{\ensuremath{t_{\text{seg}}}\xspace}

\DeclarePairedDelimiter\abs{\lvert}{\rvert}

\newcommand{\mypara}[1]{\noindent\textbf{#1}}

\newcommand{\invisible}[1]{}

\def\paperID{9}
\def\confName{3DV\xspace}
\def\confYear{2026\xspace}

\title{\ourtitle}

\author{
   Hou In Derek Pun$^{1}$ \quad Hou In Ivan Tam$^{1}$ \quad Austin T. Wang$^{1}$ \quad Xiaoliang Huo$^{1}$ \\ Angel X. Chang$^{1,2}$ \quad Manolis Savva$^{1}$\\
   $^{1}$Simon Fraser University \quad $^{2}$Alberta Machine Intelligence Institute (Amii) \\
   {\small{\url{https://3dlg-hcvc.github.io/hsm/}}}
}

\begin{document}

\newcommand{\figfirstpagefigure}{
    \vspace{-3em}
    \begin{center}
    \captionsetup{type=figure}
    \includegraphics[width=\textwidth]{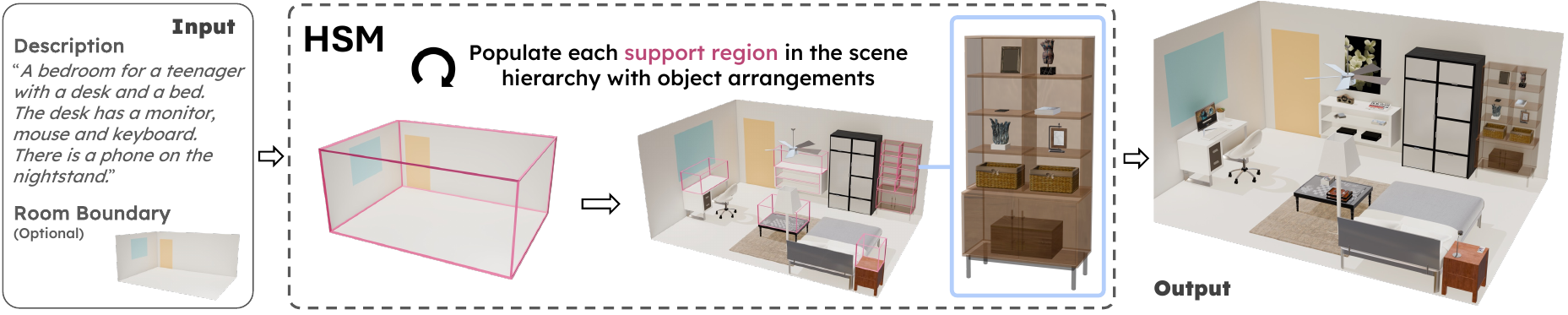}
    \vspace{-1.75em}
    \captionof{figure}{
        \textbf{Overview.}
        Given a room description and optional room boundary as input, \ours decomposes indoor scenes hierarchically and identifies valid support regions (highlighted in pink boxes) at each level of the hierarchy. The system then populates these regions by generating and optimizing object arrangements in a unified manner across scales, generating scenes with dense object arrangements.
    }
    \label{fig:teaser}
    \end{center}
    \vspace{0.5em}
}

\twocolumn[{
    \maketitle
    \figfirstpagefigure
}]

\maketitle

\begin{abstract}
Despite advances in indoor 3D scene layout generation, synthesizing scenes with dense object arrangements remains challenging.
Existing methods focus on large furniture while neglecting smaller objects, resulting in unrealistically empty scenes. 
Those that place small objects typically do not honor arrangement specifications, resulting in largely random placement not following the text description.
We present \oursfull (\ours): a hierarchical framework for indoor scene generation with dense object arrangements across spatial scales.
Indoor scenes are inherently hierarchical, with surfaces supporting objects at different scales, from large furniture on floors to smaller objects on tables and shelves.
\ours embraces this hierarchy and exploits recurring cross-scale spatial patterns to generate complex and realistic scenes in a unified manner.
Our experiments show that \ours outperforms existing methods by generating scenes that better conform to user input across room types and spatial configurations.
\end{abstract}

\section{Introduction}
\label{sec:intro}

Digital 3D indoor scenes are widely used in domains such as gaming, interior design, virtual training, and simulation for embodied AI and robotics.
Consequently, efficient and controllable generation of realistic indoor scenes has been a long-standing research problem.
While recent advances have improved scene synthesis through various approaches, most efforts have focused on arranging large furniture pieces, with less attention given to arrangement of smaller objects, such as computer peripherals and place settings.
These objects are often treated as an afterthought, placed randomly or based on predefined rules, limiting realism of the generated scenes and downstream applications.

Small objects present a unique challenge in scene generation due to their dependence on larger furniture for support.
For instance, a computer monitor is typically placed on a desk, while books are arranged on shelves.
These objects must be positioned to respect the physical constraints of their supporting furniture while adhering to user specifications.
However, existing methods rarely capture this hierarchical dependency, leading to sparse and unrealistic small object placements.
In addition, the lack of large-scale scene datasets that encompass both furniture-level and small object arrangements makes it non-trivial to train models capable of generating complex hierarchical scenes.

In this work, we introduce \oursfull (\ours), a hierarchical approach to generating indoor scenes with densely populated objects.
We employ a unified approach to placing objects, guided by both text-explicit and implicit relationships. This spans scales from room-level furniture layouts to the placement of small objects on furniture.
At each level, \ours first identifies valid \textit{support regions} (i.e., regions in which objects can be placed) and then generates compositional object arrangements within them.
This 
transforms the scene generation problem into equivalent subproblems of arranging objects on surfaces, while ensuring that the generated scene aligns with user intent.

Our key insight is that object arrangements in indoor scenes exhibit recurring spatial patterns across scales.
For instance, dining chairs surrounding a circular table and place settings arranged around a centerpiece follow a common circular pattern.
We refer to such spatial patterns as \textit{motifs}\,---\,fundamental and reusable object placement structures that are ubiquitous in indoor environments.
These motifs can be efficiently learned from a few examples and applied across different scales to generate realistic and coherent object arrangements.
By composing these motifs into \textit{scene motifs}, we enable the generation of complex arrangements across the scene hierarchy under a unified framework, facilitating structured and scalable scene generation.

We demonstrate that \ours generates realistic indoor scenes with dense object arrangements that align with user expectations and adhere to physical constraints.
In particular, our method excels at 
plausible and text-consistent
small object placement, producing coherent and detailed arrangements that enhance scene realism (e.g. neatly stacked books on shelves and well-organized stationery and accessories on desks).
We evaluate \ours against state-of-the-art scene generation methods and show that it more effectively handles complex object arrangements, generating scenes with more realistic object placements and hierarchical structures.
Our results highlight the importance of considering hierarchical relationships in indoor scenes and demonstrate that scene motifs are a powerful mechanism for generating high-quality scenes.
In summary:
\begin{itemize}
    \item We present \ours, a hierarchical framework for indoor scene generation that identifies support regions and generates structured object arrangements across different scales in a unified manner, producing realistic scenes.
    \item We introduce scene motifs, compositional structures that capture recurring spatial patterns in indoor environments, enabling the generation of complex arrangements.
    \item We show \ours produces realistic scenes with dense object arrangements that adhere to physical constraints and better align with user intent compared to prior work.
\end{itemize}

\begin{figure*}[ht]
    \includegraphics[width=\linewidth]{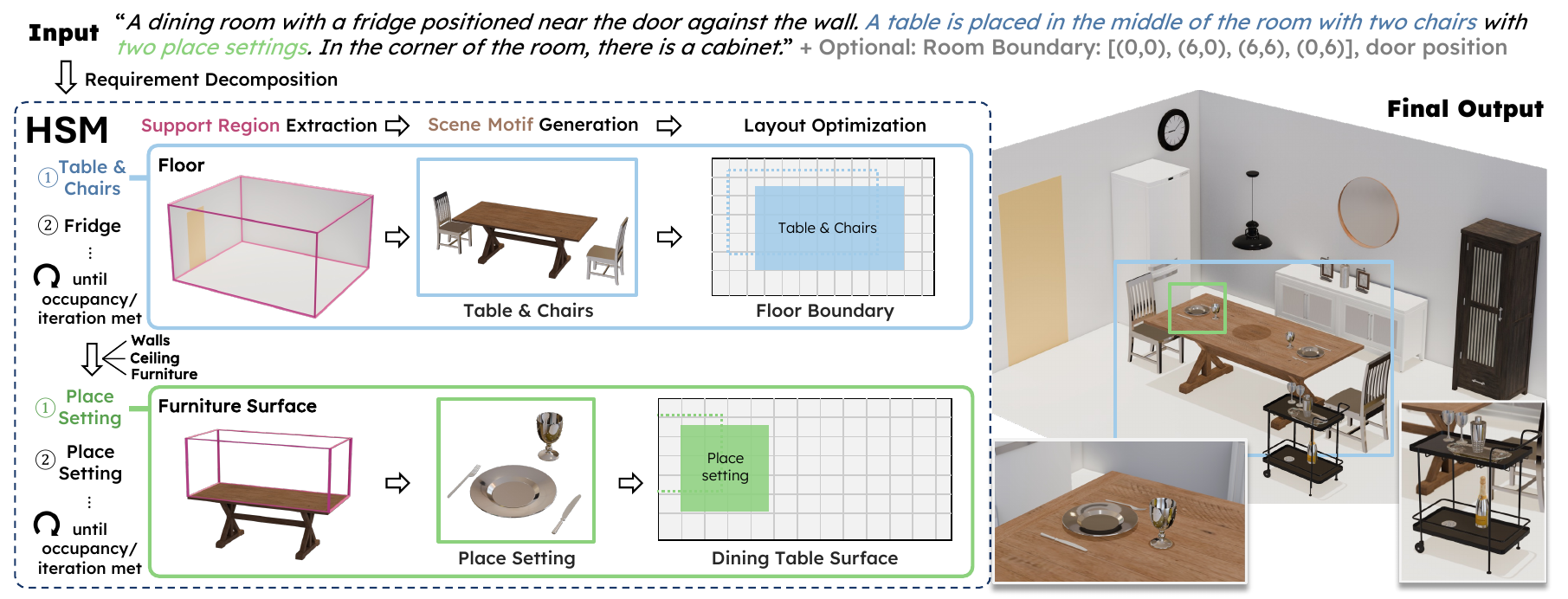}
    \vspace{-2em}
    \caption{
        \textbf{HSM framework overview.}
        Given an input text description and an optional room boundary, \ours decomposes the input into requirements at different scales, and generates the scene through a unified three-stage framework:
        1) Extract support regions for object placements;
        2) Generate appropriate scene motifs for each region;
        and 3) Optimize scene motif placements within each region.
        These steps are repeated across scales to generate a scene that aligns to the input with dense small object placements.
    }
    \vspace{-1.25em}
    \label{fig:pipeline}
\end{figure*}

\section{Related Work}
\label{sec:related}

\mypara{Indoor scene generation.}
Early approaches primarily relied on rule-based
~\cite{kjolaas2000automatic,coyne2001wordseye,xu2002constraint,merrell2011interactive,deitke2022procthor,raistrick2024infinigen}
or data-driven
~\cite{jiang2012learning,fisher2012example,chang2014learning,fisher2015activity,savva2016pigraphs,henderson2017automatic,fu2017adaptive,qi2018human,keshavarzi2020scenegen,zhang2021mageadd}
methods.
The advent of deep learning led to a shift towards learning-based methods~\cite{wang2018deep,ritchie2019fast,li2019grains,zhang2020deep,purkait2020sg,wang2021sceneformer,dhamo2021graph,paschalidou2021atiss,para2023cofs,tang2024diffuscene,lin2024instructscene,hu2024mixed,yang2024physcene,min2024funcscene,zhai2024commonscenes,zhai2024echoscene,sun2024forest2seq,wei2024planner3d,wu2025external,ye2024maan,yang2025mmgdreamer, li2024gltscene}.
These methods take various forms of input, including text descriptions, scene graphs, and images, to generate 3D scenes.
Of particular interest are methods that generate scenes from text
~\cite{tang2024diffuscene,lin2024instructscene,ye2024maan,yang2025mmgdreamer},
with recent works increasingly incorporating large language models (LLMs) into the process
~\cite{yang2024holodeck, ccelen2025design, fu2025anyhome, hu2024scenecraft, wang2024architect, littlefair2025flairgpt, yang2024llplace, wang2024chat2layout, aguina2024open, xia2024scenegenagent, wu2024diorama, sun2024layoutvlm, ma2018language, liu2025scenefunctioner, sun2025hierarchically, liu2025worldcraft, deng2025global, dong2025hiscene, feng2025casagpt, ling2025scenethesis, bai2025freescene, su2025chord}.
While these methods enable text-conditioned scene generation, most focus solely on arranging large furniture, neglecting the small objects that are ubiquitous in indoor environments.
More recent works have begun integrating small objects into the scene generation process
~\cite{yang2024holodeck, fu2025anyhome, wu2024diorama, wang2024architect, littlefair2025flairgpt, zhang2019active, zhang2023automatic, liu2025worldcraft, dong2025hiscene, ling2025scenethesis, pfaff2025steerable, liu2025agentic, ran2025direct, huang2025video, zhou2025roomcraft, gu2025artiscene}.
Additionally, some methods attempt to place small objects in existing scenes
~\cite{majerowicz2013filling, yu2015clutterpalette, huang2025fireplace, abdelreheem2025placeit3d}.
However, these methods often treat small objects in a simplified or specialized manner, such as using random placement. This limits both the diversity and controllability of their arrangements.
Our work introduces a unified hierarchical framework for scene generation that conditions object placement on precise language descriptions, capturing object relationships at all scales.

\mypara{Hierarchical scene generation.}
Exploiting the hierarchical nature of scene generation has been studied for many years~\cite{li2019grains, wang2019planit, gao2023scenehgn, yu2011make, littlefair2025flairgpt, min2024funcscene, ye2024maan, sun2025hierarchically, dong2025hiscene, su2025chord}. 
Among recent works, Architect~\cite{wang2024architect} uses a parallel approach to generate large and small objects but relies on 2D inpainting and thus suffers from 3D inconsistencies.
Furthermore, Architect does not generate scenes with precise control over object arrangements.
SceneFunctioner~\cite{liu2025scenefunctioner} groups objects and parses relationships within each group, similar to \ours, but their approach realizes relationships using LLM-predicted anchor rules rather than recurring motifs and does not handle small objects. 
While most such papers primarily capture the semantic hierarchy of objects in scenes, we further capture the repetition of object relationships at each scale, in that small objects can be functionally and spatially arranged equivalently to large objects.

\mypara{Support region prediction.}
To place small objects in scenes, it is first beneficial to determine suitable placement regions.
Various methods have been proposed to predict support surfaces in indoor environments from 2D images~\cite{wu2024diorama,rozumnyi2023estimating,stekovic2020general,guo2013support,ren20183d,ge20243d}.
While these methods are effective, extending them to synthetic 3D scenes is non-trivial, as they rely on sight lines to surface geometry, necessitating camera viewpoint selection and introducing occlusion issues.
In contrast, our approach employs geometric reasoning-based support region extraction to identify valid support surfaces on objects, enabling dense and precise small object arrangements in generated scenes.

\section{Method}
\label{sec:methods}

Given a text description \inputdesc of an indoor scene and an optional room boundary as a list of vertices as input, \ours generates the scene iteratively through a unified hierarchical framework, as shown in \cref{fig:pipeline}.
It first uses a vision language models (VLM) to extract a room type and decompose \inputdesc into a list of required objects at each scale (\cref{sec:methods/requirement}).
The scene is then constructed through three key steps at each hierarchical level: support region extraction (\cref{sec:methods/support_region}), scene motif generation (\cref{sec:methods/scenemotif}), and layout optimization (\cref{sec:methods/placement}).
These steps are iteratively applied to an initially empty scene, first placing room-level furniture and then arranging small objects on furniture surfaces. 
After placing the initial objects, if the occupancy is below a threshold $\occthresh$ and the number of iterations below $\iterthresh$, a VLM is prompted for additional objects to add to the scene beyond the input description, conditioned based on the predicted room type. 
Throughout this process, \ours leverages a library of learned motifs to compose complex spatial relationships as scene motifs (\cref{sec:methods/motif}), which are instantiated to produce physically valid object arrangements in the scene.

\subsection{Scene Motifs}
\label{sec:methods/motif}

Inspired by SceneMotifCoder (SMC)~\cite{tam2024scenemotifcoder}, we define a \emph{motif} as an atomic spatial pattern between objects that can be extracted from a few examples and used as a template for generating new arrangements.
For instance, by analyzing arrangements such as ``a stack of books'' and ``a stack of plates'', we can learn a \texttt{stack} motif that captures the vertical alignment pattern and apply it to new objects to generate novel arrangements.
Such motifs are applicable across different scales, from room-scale furniture to small objects.

We extend this concept to capture more complex arrangement structures through \emph{scene motifs}.
A scene motif is a composition of one or more learned motifs that represents a set of spatial relationships between objects.
For example, a scene motif for a desk setup may consist of a \texttt{stack} motif for the books on the desk, a \texttt{left\_of} motif for placing a lamp to the left of the books, and an \texttt{in\_front} motif for positioning a laptop in front of the books.
By leveraging these compositions, scene motifs dynamically capture a broader range of spatial relationships, enabling the generation of more complex and diverse object arrangements.
We describe the generation process of scene motifs in \cref{sec:methods/scenemotif}.

As in SMC, motifs are implemented as visual programs that encode spatial relationships between objects using programmatic constructs.
Each motif is defined as a Python function that takes specific arguments to generate object arrangements when executed.
For example, a \texttt{stack} motif may be implemented as a function that iteratively stacks objects vertically using a \texttt{for} loop, with parameters for count and object type (e.g., \texttt{stack(4, book)} to generate ``a stack of four books'').
\ours assumes a pre-learned library of motifs exists, and uses it to dynamically composes scene motifs.
See \cref{sec:supp/implementation/motif_library} for more details.

\begin{figure}
    \centering
    \includegraphics[width=\linewidth]{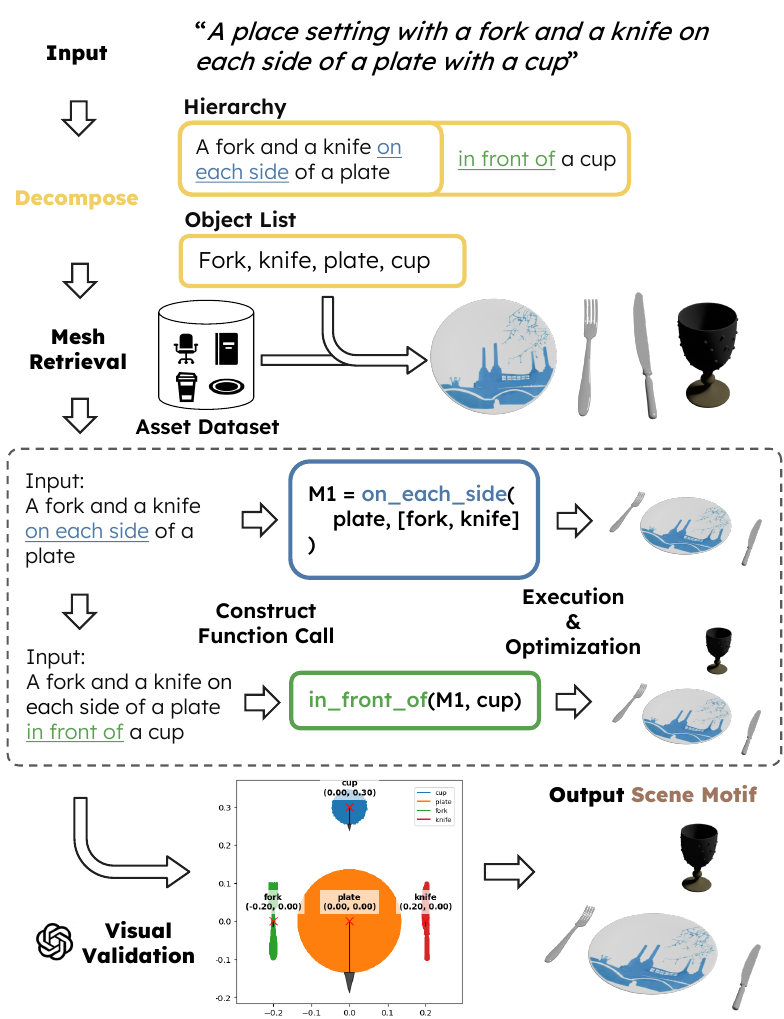}
    \vspace{-1.75em}
    \caption{
        \textbf{Scene motif generation process.}
        An input description is first decomposed into a hierarchy of motifs.
        We then retrieve the corresponding 3D assets and generate the scene motifs iteratively, starting from the innermost motif (\texttt{on\_each\_side}) and expanding to the outermost motif of the hierarchy (\texttt{in\_front\_of}).
        The generated scene motif is visually validated with a VLM.%
    }
    \label{fig:motif}
    \vspace{-1em}
\end{figure}

\subsection{Input Description Requirement Decomposition}
\label{sec:methods/requirement}

Given the input text description, \ours first uses a VLM to extract the room type and an initial list of objects to include in the scene. 
Objects are grouped by the VLM according to their supporting architectural elements or objects.
This breaks the generation task into smaller, more manageable subproblems that can be addressed within a unified framework while minimizing the risk of losing details in subsequent steps (see \cref{fig:pipeline}).
For each object, we extract its quantity, appearance, and dimensions from the input description, inferring plausible values when unspecified.
If the input lacks a room boundary, we prompt the VLM to generate one including door, window placement and room height.
See \cref{sec:supp/prompts} for the VLM prompts used.

\subsection{Support Region Extraction}
\label{sec:methods/support_region}

\begin{figure*}
    \centering
    \includegraphics[width=\linewidth]{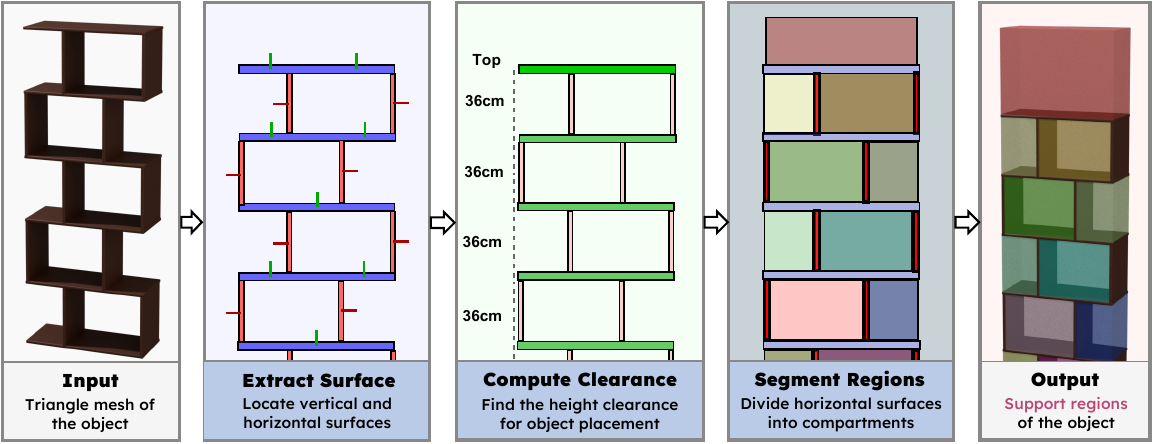}
    \vspace{-1.5em}
    \caption{
        \textbf{Support region extraction.}
        Given a triangle mesh of the object, such as the example shelf unit, we first extract vertical and horizontal surfaces.
        We then compute the height clearance for each horizontal surface and use the vertical surfaces to segment them into compartments.
        The result is a set of support regions that can be populated with objects.
    }
    \label{fig:support_extraction}
    \vspace{-1em}
\end{figure*}

To place objects, we first identify suitable placement areas, referred to as \textit{support regions} $S$.
Support regions are surfaces capable of accommodating object placement.
At the room level, these include the floor, walls, and ceiling, while at the furniture level, these correspond to horizontal surfaces such as tabletops and shelves.
We parametrize each $s_i\in S$ by a sub-mesh and height clearance $h_i$.
Wall support regions are extracted by projecting scene motifs within \wallprojthresh onto the wall to exclude blocked areas.

While room-level support regions can be 
extracted from the room boundary,
furniture-level support regions require more complex analysis due to the geometric intricacies of furniture surfaces (e.g., multi-level shelves).
We first identify horizontal and vertical surfaces on a furniture mesh with $n$ vertices and $p$ triangle faces $F \in \{1, \ldots, n\}^{p\times 3}$, parameterizing each surface as a mesh subset of $F$.
Inspired by \citep{karpathy2013object}, we cluster $F$ into planar surfaces by seeding each cluster $c_j$ with the largest unclustered $f_{j_0}$ by area and adding $f_i$ to $c_j = \{j_k\mid k=1,\ldots, \abs{c_j}\}$ if its normal is in the same direction as the normals of the first triangle and adjacent triangle. 
Formally, we check if $n_i \cdot n_{j_0} \geq \normalthresh$ and $n_i \cdot n_{j_{\text{adj}}} \geq \adjacentthresh$, where $n_i$ is the normal of $f_i$ and $f_{j_{\text{adj}}}$ shares an edge with $f_i$. While building up a cluster, we traverse adjacent faces $f_i$ by area (from largest to smallest).

We identify horizontal and vertical surfaces by fitting a plane to each $c_i$ and thresholding the vertical component of the normal.
To ensure functional utility, we compute $h_i$ between horizontal surfaces and discard those which have low clearance ($h_i < \clearancethresh$).  %
For top surfaces without a ceiling, we assign a default clearance of \topheight, in order to put a reasonable limit on the size of small objects placed on top of furniture.
Finally, horizontal surfaces are split by intersecting vertical surfaces thicker than \segthresh, ensuring continuity within each region.
\cref{fig:support_extraction} illustrates this process for a shelf unit.
See \cref{sec:supp/implementation/support_region} for implementation details.

\subsection{Scene Motif Generation}
\label{sec:methods/scenemotif}

We populate $S$ by generating \textit{scene motifs} $M$.
Given the objects $O_i$ selected for each $m_i \in M$ and the overall room description, \ours prompts a VLM to produce a natural language description of the
object layout.
The VLM then uses this description to group objects into $M$ based on spatial and functional relationships.
For example, given the description, ``a room with six chairs around a table and a sofa facing a TV,'' in the floor support region, the system groups the chairs around the table and the sofa facing the TV.

Each $O_i$ is decomposed into a sequence of one or more motifs $m_{ik}$ that capture the spatial relationships between objects.
We first use a VLM to identify a primary motif that serves as the base object arrangement, relative to which the rest of the objects in $O_i$ can be placed.
\cref{fig:motif} demonstrates a place setting example, in which the utensil-plate arrangement forms the primary motif, with the cup positioned relative to it.
The VLM then predicts an arrangement of the remaining objects around this primary motif based on their spatial and functional relationships.
The result is a hierarchical sequence of motifs that define the object arrangements within $m_i$, validated by the VLM against the input description to ensure completeness and coherence.
If validation fails, the reason is included in retry attempts.

The scene motif is instantiated by creating the primary motif and progressively traversing the hierarchy to model additional objects.
We first retrieve a suitable 3D mesh for each object from a database based on its category, appearance description, and dimensions (see \cref{sec:supp/implementation/retrieval} for details on the retrieval process).
For each motif, we use a VLM to select a matching program from a pre-learned library of motifs and generate an appropriate function call to instantiate it.
This function is then executed to generate an object arrangement using the retrieved meshes.
The resulting arrangement is subsequently used as input for the next motif in the hierarchy, iteratively constructing the scene motif until all objects are placed. 
To ensure physical plausibility between objects, we apply the same spatial optimization used in SMC---resolve collisions, move objects closer, and simulate gravity---while respecting the hierarchy.
Motifs are treated as single units during optimization.

To ensure the generated scene motif aligns with the input description, we employ a verification process after its generation. We generate top-down and front orthographic projections of the scene motif and prompt a VLM to validate the arrangement against the input description based on these projections. If validation fails, the failure reason is used to guide the VLM in retrying the generation process.

\subsection{Layout Optimization}
\label{sec:methods/placement}

The final step in populating a support region is placing the generated scene motifs within it.
This is achieved through a three-stage process.
In the first stage, we provide an orthogonal projection of the support region, the bounding boxes of the scene motifs, and the sub-scene description to a VLM, prompting it to suggest initial placements and determine whether the scene motifs should be wall-aligned.
The VLM is instructed to consider both explicit constraints from the input description and implicit constraints derived from common arrangements and usage patterns to generate a reasonable initial layout.
The second stage refines the layout using an optimization solver.
Inspired by Holodeck~\cite{yang2024holodeck}, we employ a grid-based depth-first search (DFS) solver.
Starting with the scene motif that has the largest footprint, the solver iteratively refines placements by enforcing the following constraints:
1) scene motifs must be placed within the support region,
2) wall-aligned scene motifs must have their back against a wall and face into the room, and
3) there should be no overlap between scene motifs.
The solver returns the first valid layout or otherwise the initial placement if a set time limit is exceeded.
Finally, we apply a scene-level spatial optimizer to refine placements by eliminating mesh collisions and ensuring valid support.
For each scene motif, we resolve collisions and support issues using relationship-aware rules and raycasting, making minimal adjustments while preserving hierarchy and layout. 
We provide the details for the layout optimization in \cref{sec:supp/implementation/layout_optimization}.

These three steps\,---\,support region extraction, scene motif generation, and layout optimization\,---\,are repeated at each hierarchy level, from room-scale furniture arrangements to fine-grained small object placements on furniture surfaces.
This unified framework ensures a structured and efficient generation process, preserving spatial coherence and physical validity across scales to produce complex indoor scenes with realistic object arrangements.

\section{Experimental Setup}
\label{sec:experiments}

\begin{figure*}
    \centering
    \includegraphics[width=\linewidth]{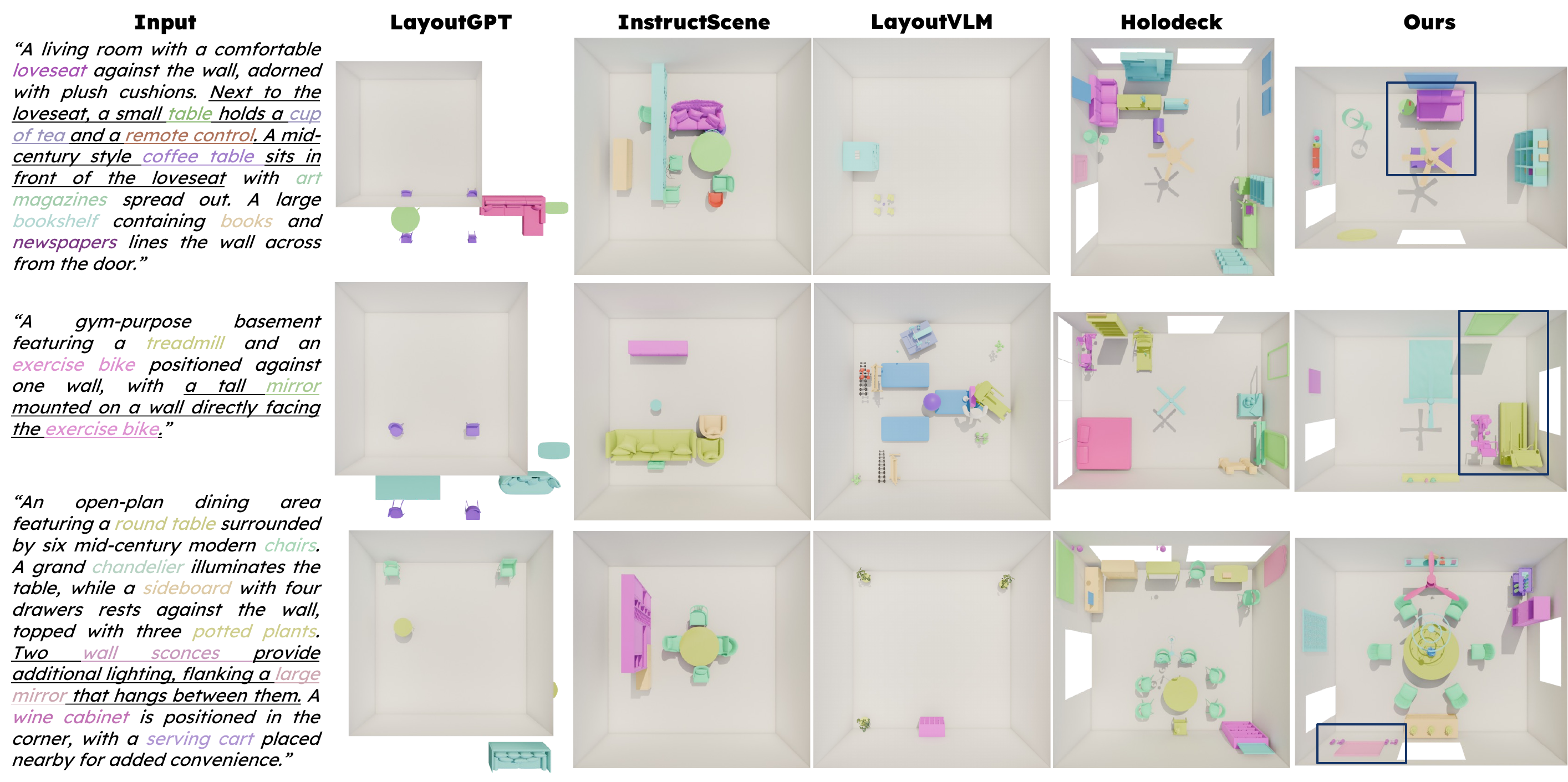}
    \vspace{-2em}
    \caption{
        \textbf{Qualitative comparisons at the scene level.}
        Objects and spatial relationships in the input text are highlighted with colors and underlines, and spatial relationships are emphasized using boxes.
        \ours produces more coherent spatial arrangements and is better aligned to the input compared to existing approaches.
    }
    \label{fig:qualitative}
    \vspace{-1em}
\end{figure*}

\subsection{Scene Generation}
\label{sec:experiments/scene_generation}

We evaluate \ours on the task of text-conditioned indoor scene generation using 3D assets from the Habitat Synthetic Scenes Dataset (HSSD-200)~\cite{khanna2023hssd}.
We select HSSD as it comprises 211 synthetic indoor scenes with a diverse collection of high-quality 3D assets --- both furniture and small objects --- making it well-suited for learning motifs and retrieving objects.
We use \textit{gpt-4o-2024-08-06}~\cite{achiam2023gpt} for all VLM usage. 
For our experiments, we set $\occthresh=0.3$ and $\iterthresh=2$ for floor support region, with $\occthresh=0.5$ and $\iterthresh=1$ for the others, $\wallprojthresh=\SI{1.5}{\metre}$ for wall projection.

\mypara{Baselines.}
We compare \ours against four recent scene generation methods: LayoutGPT~\cite{feng2024layoutgpt}, InstructScene~\cite{lin2024instructscene}, LayoutVLM~\cite{sun2024layoutvlm} and Holodeck~\cite{yang2024holodeck}.
LayoutGPT is the first to leverage an LLM for indoor scene generation.
It uses scenes from the 3D-FRONT dataset~\cite{fu20213dfront,fu20213dfuture} as in-context examples and prompts an LLM to generate layouts in CSS format.
InstructScene employs a graph diffusion model with a semantic scene graph to generate 3D object layouts from text descriptions.
LayoutVLM is a framework that uses VLM with differentiable optimization to generate 3D object layouts from text descriptions.
Holodeck is a comprehensive system that integrates LLM-based generation with optimization steps to produce room boundaries and object placements for embodied AI simulations.
We follow the original implementations of these methods and use their respective object databases for object retrieval: 3D-FRONT~\cite{fu20213dfront,fu20213dfuture} for LayoutGPT and InstructScene, and Objaverse~\cite{deitke2023objaverse} for LayoutVLM and Holodeck.
Since only Holodeck can generate architectural elements,
we provide a standardized 6 m × 6 m square floor plan with walls as input for the other methods to ensure fair comparison.

\mypara{Input Text Descriptions.}
For all methods, we use the first 100 text descriptions from SceneEval-500, a dataset introduced by SceneEval~\cite{tam2025sceneeval}, a recent framework for evaluating text-to-indoor scene generation methods.
The descriptions vary in complexity (40 easy, 40 medium, 20 hard) and encompass a wide range of scene types and object arrangements.
Each scene description is paired with human-annotated ground truth scene properties (e.g., object counts, attributes, and spatial relationships) to assess how well the text match the generated scenes.

\begin{table*}
\centering
\resizebox{\linewidth}{!}
{
\begin{tabular}{@{} l rrr rrrr rrrrrr r r@{}}
\toprule

& \multicolumn{3}{c}{Text-Image Score} 
& \multicolumn{4}{c}{SceneEval Fidelity} 
& \multicolumn{6}{c}{SceneEval Plausbility} 
& \multirow{2.5}{*}{\shortstack{Avg.\ $\#$Obj\\ per Scene}}
\\ 
  
\cmidrule(l){2-4} \cmidrule(l){5-8} \cmidrule(lr){9-14} 
& $\uparrow~$BLIP & $\uparrow~$CLIP  & $\uparrow~$VQA

& $\uparrow~$CNT$_{\%}$ & $\uparrow~$ATR$_{\%}$ & $\uparrow~$OOR$_{\%}$ & $\uparrow~$OAR$_{\%}$

& $\downarrow~$COL$_{ob\%}$ & $\downarrow~$COL$_{sc\%}$ & $\uparrow~$SUP$_{\%}$ & $\uparrow~$NAV$_{\%}$
& $\uparrow~$ACC$_{\%}$ & $\downarrow~$OOB$_{\%}$

\\
\midrule

LayoutGPT
& 0.0613 & 0.1670 & 0.2964
& 19.54 & 18.98 &  2.87 &   5.24
& \textbf{12.96} & \textbf{30.00} & 28.24 & \textbf{100.00}
& 47.29 & 73.11 
& 5.17
\\

InstructScene                           
& 0.0845 & 0.1681 & 0.4082
& 25.48 & 22.26 & 11.17 & 10.48
& 51.18 & 84.00 & \underline{75.09} & \underline{99.53}
& 77.30 & 22.92
& 8.07
\\
\midrule

LayoutVLM
& 0.0857 & 0.1612 & 0.3268
& 41.19 & 22.26 & 8.60 & 23.29
& 36.09 & 69.00 & 67.96 & 98.75
& 85.91 & 4.14
& 11.36
\\

Holodeck
& \underline{0.1230} & \underline{0.1820} & \underline{0.5549}
& \underline{44.64} & \underline{39.42} & \underline{20.92} & \underline{49.60}
& 17.32 & 73.00 & 62.12 & 99.45
& \textbf{90.55} & \textbf{1.30}
& 24.71
\\

\oursshort (ours)
& \textbf{0.1748} & \textbf{0.1841} & \textbf{0.5627}
& \textbf{61.30} & \textbf{59.49} & \textbf{40.40} & \textbf{70.28}
& \underline{16.42} & \underline{61.00} & \textbf{85.44} & 98.97
& \underline{86.80} & \underline{2.13}
& 20.65
\\

\bottomrule
\end{tabular}
}
\vspace{-0.75em}
\caption{
\textbf{Evaluation with text-image scores and SceneEval metrics.}
Overall, \ours outperforms prior work along metrics measuring the fidelity of object placements relative to the input text (Text-Image Score and SceneEval Fidelity metrics). 
For plausibility metrics, \ours has the highest support rates. 
LayoutGPT and InstructScene have better collision and navigation but that is due to placing far fewer objects (last column), most of which are out of the scene (OOB).
Bold indicates highest results, underlined denotes second highest.
}
\vspace{-0.5em}
\label{tab:scene_quantitative}
\end{table*}

\mypara{Metrics.}
We use four fidelity metrics from SceneEval to assess how well the generated scenes align with the input text description: Object Count (\textbf{CNT}), Object Attribute (\textbf{ATR}), Object-Object Relationship (\textbf{OOR}), and Object-Architecture Relationship (\textbf{OAR}), using annotations from SceneEval-100 as ground truth. 
To evaluate object placement, we use five plausibility metrics capturing implicit human expectations: Object Collision (\textbf{COL}), Object Support (\textbf{SUP}), Scene Navigability (\textbf{NAV}), Object Accessibility (\textbf{ACC}), and Object Out-of-Bound (\textbf{OOB}).

We also conduct a user perceptual study to compare generated scenes from \ours and Holodeck.
We select Holodeck as the baseline for comparison due to its strong generation performance and ability to place small objects.
The study consists of two parts: scene-level evaluation and small object-level evaluation.
At the scene level, we randomly select 25 scene pairs generated by each method for the same input description.
Given top-down renderings, participants assess which scene better aligns with the input text descriptions (\textbf{Fidelity}) and exhibits more physically realistic object placements (\textbf{Plausibility}).
At the small object level, we randomly select 30 pairs of furniture populated with small objects and provide close-up renderings.
Each pair originates from the same scene description, and participants evaluate the placements based on fidelity and plausibility.
The study was conducted with 25 participants, and we report the percentage of participant preferences for each method across both evaluation criteria and levels.
See \cref{sec:supp/user_study} for the user study instructions.

We also evaluate alignment between generated scenes and input descriptions by rendering top-down views and computing text-image similarity using BLIP-2~\cite{li2023blip}, Long-CLIP~\cite{zhang2024longclip} and VQAScore~\cite{lin2024evaluating}. 
Similarly, we report alignment between small object arrangements and the corresponding text descriptions using the same metrics and the set of 30 populated furniture from the user study. 
For methods that do not generate small objects, we select the best-matching furniture available for evaluation.
The score is set to $0$ if no matching furniture of the same category is found.

\subsection{Support Region Extraction}
\label{sec:experiments/support_region}

As support region extraction from 3D furniture meshes is a key component of our approach and an essential step for placing small objects, we evaluate this step in isolation.
To this end, we manually annotate 100 furniture items (e.g., tables, shelves, sofas) from the HSSD-200 dataset~\cite{khanna2023hssd} with ground truth support regions.
Each annotated region is represented by a set of faces defining the support surface and its height clearance.
A single furniture piece may contain multiple support regions of varying shapes and sizes (e.g., as in \cref{fig:support_extraction}).
The 100 objects we annotated have a total of 529 regions, with an average of five regions per object.
See the~\cref{sec:supp/support_region_dataset} for details on the annotation process.

\mypara{Metrics.}
We evaluate the extracted support regions against ground truth annotations using two metrics: Intersection over Union (IoU) of the regions' volume, measuring the overlap between predicted and ground truth support regions; and F1-score, which provides a balanced measure between precision (accuracy of predicted regions) and recall (completeness of detected regions) at a threshold of 0.5.

For IoU computation, each ground truth and predicted support surface is projected onto the horizontal plane
to simplify calculation of overlapping areas.
Subsequently, the intersection volume of the two regions is determined by multiplying the area of intersection of the projected surfaces with the overlap of their heights along the vertical axis.
To emphasize the importance of support surface alignment, we apply a height threshold $t_d$, when computing the IoU. 
If the vertical distance between ground truth and predicted surfaces exceeds $t_d$, the IoU is set to $0$.
We fix $t_d$ to $10$ cm to ensure strict correspondence in support surface height alignment.
The Hungarian algorithm is applied to these IoU values to determine optimal region-region correspondence, and match predicted and actual support surfaces.

\mypara{Baselines}.
We compare our support region extraction approach against a baseline which only predicts support regions on the top surface of an object, thus matching Holodeck~\cite{yang2024holodeck}'s ray casting approach for object placement.

\section{Results}
\label{sec:results}

\begin{table}
\centering
\resizebox{0.8\linewidth}{!}{
\begin{tabular}{@{} l l cc r@{}}
\toprule
Level & Method
& $\uparrow~$Fidelity\textsubscript{\%} & $\uparrow~$Plausibility\textsubscript{\%} \\

\midrule

\multirow{2}{*}{Scene level}
& Holodeck          & 23.36          & 30.24 \\
& \oursshort (ours) & \textbf{76.64} & \textbf{69.76} \\

\midrule
\multirow{2}{*}{Small object level}
& Holodeck          & 18.40          & 26.80 \\
& \oursshort (ours) & \textbf{81.60} & \textbf{73.20} \\

\bottomrule
\end{tabular}
}
\vspace{-0.5em}
\caption{
\textbf{User study results.}
Holodeck~\cite{yang2024holodeck} compared to \ours at scene and small object levels.
\ours is preferred at both levels.
}
\label{tab:user_study}
\end{table}

\begin{table}
\centering
\resizebox{0.7\linewidth}{!}{
\begin{tabular}{@{} l rrr r@{}}
\toprule
& $\uparrow~$BLIP & $\uparrow~$CLIP  & $\uparrow~$VQA \\
\midrule

LayoutGPT & 0.0742 & 0.0144 & 0.1246 \\
InstructScene & 0.1129 & 0.0413 & 0.1272 \\
LayoutVLM & 0.1432 & 0.0808 & 0.1402 \\
Holodeck & 0.2183 & 0.1250 & 0.1702 \\
\oursshort (ours) & \textbf{0.2497} & \textbf{0.1582} & \textbf{0.1732} \\

\bottomrule
\end{tabular}
}
\vspace{-0.5em}
\caption{
\textbf{Evaluation on small object placements.}
\ours outperforms prior work across all metrics. 
}
\label{tab:small_object_quantitative}
\vspace{-0.5em}
\end{table}

\begin{figure}
    \centering
    \includegraphics[width=\linewidth]{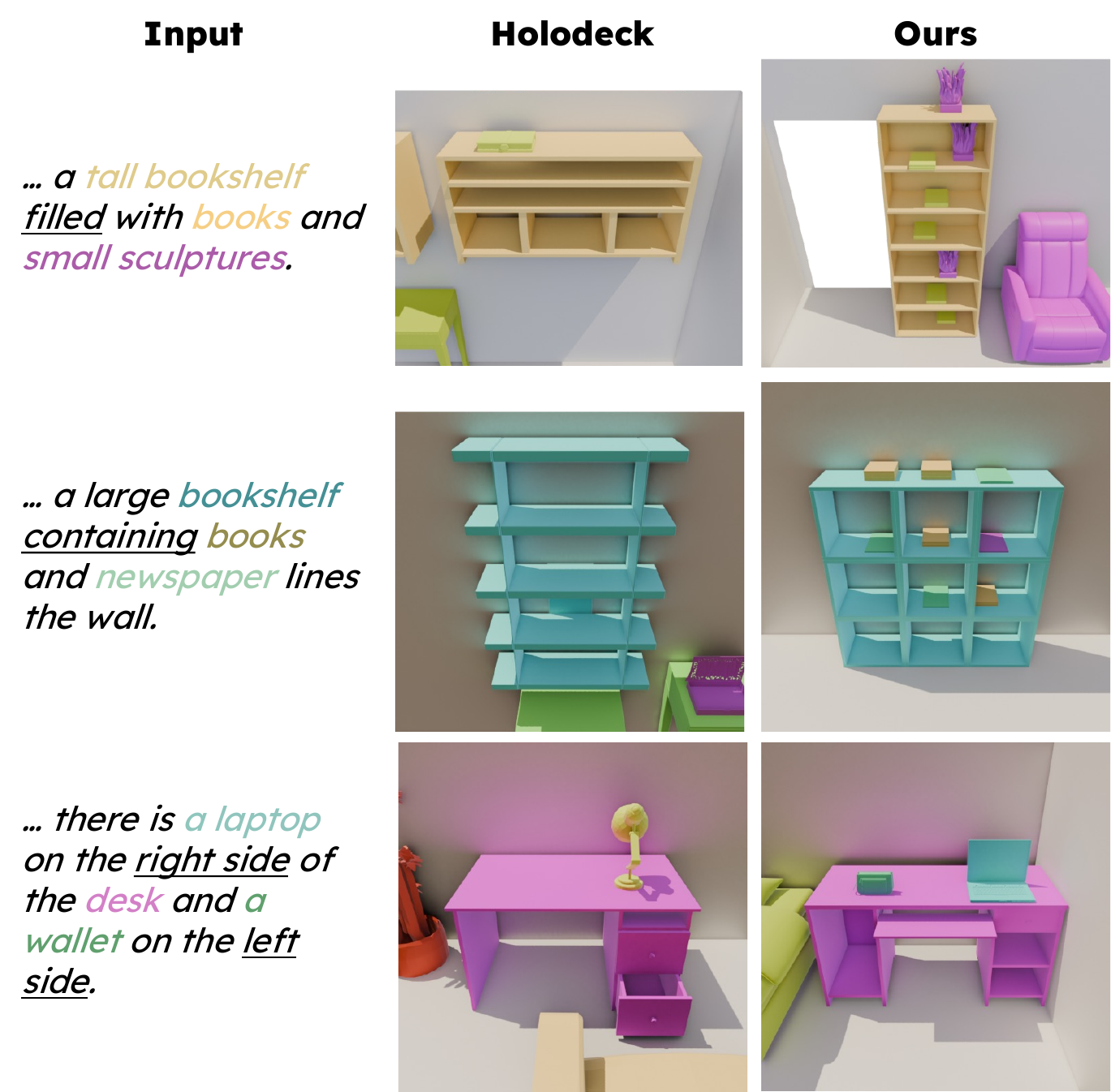}
    \vspace{-1.5em}
    \caption{
        \textbf{Qualitative comparison of object placements.}
        Each row shows close-up views of object arrangements from the input description.
        \ours better follows the spatial relationships and placement instructions specified in the input text.
    }
    \label{fig:small_object_qualitative}
    \vspace{-0.5em}
\end{figure}

\begin{figure}
    \centering
    \includegraphics[width=\linewidth]{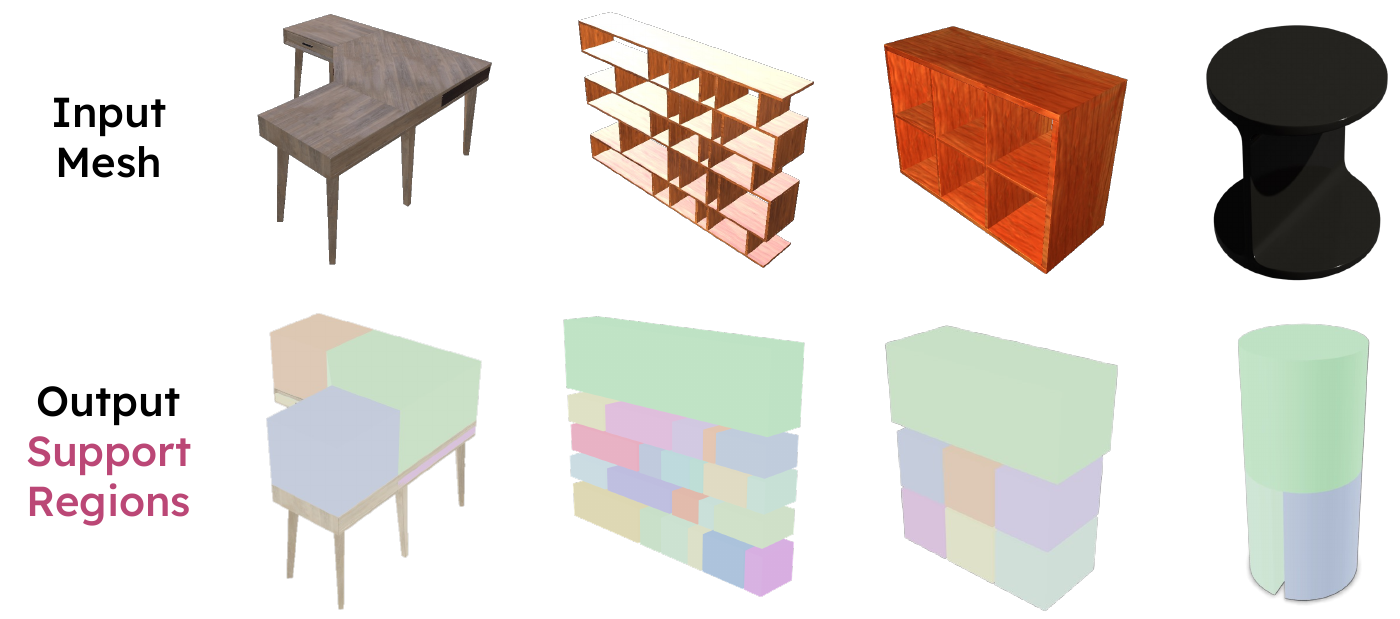}
    \vspace{-1.75em}
    \caption{
        \textbf{Support region extraction examples.}
        Colored boxes are extracted support regions for accurately placing smaller items.
    }
    \label{fig:support_example}
    \vspace{-1.25em}
\end{figure}

\begin{table*}
\centering
\resizebox{\linewidth}{!}
{
\begin{tabular}{@{} l rrrr rrrrrr r r@{}}
\toprule

& \multicolumn{4}{c}{SceneEval Fidelity} 
& \multicolumn{6}{c}{SceneEval Plausbility} 
& \multirow{2.5}{*}{\shortstack{Avg.\ $\#$Obj\\ per Scene}}
\\ 
  
\cmidrule(l){2-5} \cmidrule(lr){6-11} 
 
& $\uparrow~$CNT$_{\%}$ & $\uparrow~$ATR$_{\%}$ & $\uparrow~$OOR$_{\%}$ & $\uparrow~$OAR$_{\%}$
& $\downarrow~$COL$_{ob\%}$ & $\downarrow~$COL$_{sc\%}$ & $\uparrow~$SUP$_{\%}$ & $\uparrow~$NAV$_{\%}$
& $\uparrow~$ACC$_{\%}$ & $\downarrow~$OOB$_{\%}$
\\
\midrule

\oursshort (ours)
& \textbf{61.30} & \underline{59.49} & \textbf{40.40} & \textbf{70.28}
& \textbf{16.42} & \textbf{61.00} & \underline{85.44} & \textbf{98.97}
& \underline{86.80} & \textbf{2.13}
& 20.65
\\

- w/o scene motifs 
& 54.79 & 53.28 & \underline{32.66} & 50.20
& \underline{24.12} & 87.00 & \textbf{87.40} & 97.15
& 84.78 & \underline{2.26}
& 28.81
\\

- w/o scene spatial optimizer
& \underline{55.94} & \textbf{60.95} & 30.66 & \underline{61.85}
& 25.73 & \underline{74.00} & 83.98 & 98.37
& \textbf{87.70} & 2.95
& 22.04
\\

- w/o DFS solver
& 52.87 & 49.64 & 26.36 & 54.22
& 28.96 & 75.00 & 75.83 & \underline{98.78}
& 73.34 & 12.76
& 19.20
\\

\bottomrule
\end{tabular}
}
\vspace{-0.75em}
\caption{
\textbf{Ablation study.}
Ablations of \ours generate scenes with lower fidelity.
Removing scene motifs reduces fidelity, as the VLM must handle individual object placement rather than leveraging grouped structures.
Disabling the spatial optimizer reduces plausibility with higher COL.
Disabling the DFS solver causes the largest drop across most metrics with lower fidelity and reduced plausibility 
due to invalid object placements outside support regions.
Bold indicates highest results, underlined denotes second highest.
}
\label{tab:ablation}
\vspace{-1.2em}
\end{table*}

\subsection{Scene Generation}
\label{sec:results/scene_generation}

\cref{fig:qualitative} presents generated scenes from \ours and the baselines, while \cref{tab:scene_quantitative} reports the quantitative evaluation results.
\ours outperforms the baselines as measured by text-image score and SceneEval fidelity metrics, demonstrating better alignment between generated scenes and input descriptions.
Qualitative results further highlight \ours’s ability to capture precise requirements specified in the input.
For example, \ours is the only method that accurately generates a small table next to a loveseat with a coffee table in front in the first row, and two wall sconces flanking a mirror in the third row.
In contrast, LayoutGPT frequently places objects outside the room. InstructScene and LayoutVLM also place objects outside room boundaries, and exhibit limited object variety and layout diversity.
Additional analyses are in the supplement: computational cost and runtime breakdown (\cref{sec:supp/implementation/computation}), breakdown of results by difficulty (\cref{tab:supp/scene_quantitave_difficulty}), and evaluation of \ours using an open-source VLM (\cref{sec:supp/implementation/open_source_vlm}).

\cref{tab:small_object_quantitative} shows that \ours outperform prior work on small object placements.
Results from our user study (\cref{tab:user_study}) further demonstrate that \ours is preferred at both the scene and small object levels in fidelity and plausibility.
While Holodeck can generate small objects, it only considers the top surfaces of furniture items, leaving interiors unrealistically empty (as shown in \cref{fig:small_object_qualitative}).
In contrast, \ourspossessive support region analysis identifies valid placement regions across all object surfaces, enabling denser and more realistic small object arrangements.
By leveraging a hierarchical approach, \ours produces more structured and realistic object placements across all scales within a unified framework. See~\cref{fig:appendix_qualitative} and ~\cref{fig:appendix_render} for more qualitative results.

\subsection{Support Region Extraction}
\label{sec:results/support_region}

\cref{fig:support_example} shows examples of extracted support regions.
\ours achieves an average IoU of $60.27\%$ and F1-score of $48.54\%$ against ground truth annotations.
In contrast, the baseline that considers only top surfaces performs significantly worse, with an IoU of $32.54\%$ and an F1-score of $16.80\%$.

This demonstrates that \ours more effectively identifies valid support regions within furniture items, which is essential for accurate small object placement in scenes.

\subsection{Ablation}
\label{sec:results/ablation}

To evaluate the contribution of individual components, we ablate key elements of \oursshort:
1) \textit{w/o scene motif}\,---\, removing scene motifs and treating all objects as individual 
pieces,
2) \textit{w/o scene spatial optimizer}\,---\, removing scene spatial optimization and directly using DFS solver positions to place scene motifs, and 
3) \textit{w/o DFS solver}\,---\, removing DFS solver and directly using VLM-provided positions to place scene motifs.
\cref{tab:ablation} shows that removing any component results in worse performance.
See~\cref{sec:supp/implementation/quantitative} for a detailed analysis and qualitative comparison.

\section{Conclusion}
\label{sec:conclusion}

We presented \oursfull(\ours), a hierarchical framework for indoor 3D scene generation that produces dense object arrangements across spatial scales.
Our approach models indoor scenes as a hierarchy of support regions, each to be populated with objects.
By leveraging scene motifs, \ours generates object arrangements at multiple scales, from room-level furniture layouts to fine-grained small object placements, exploiting recurring spatial patterns.
We believe \ourspossessive unified hierarchical framework represents a significant step toward generating densely populated and realistic indoor environments.

\vspace{4pt}
\noindent\textbf{Acknowledgments.}
This work was funded in part by the Sony Research Award Program, a CIFAR AI Chair, a Canada Research Chair, NSERC DG, and enabled by support from \href{https://alliancecan.ca/}{Digital Research Alliance}.
We thank Jiayi Liu, Weikun Peng, and Qirui Wu for helpful discussions.

{
    \small
    \bibliographystyle{ieeenat_fullname}
    \bibliography{main}
}

\clearpage
\clearpage
\maketitlesupplementary
\appendix

\begin{figure*}
    \vspace{5em}
    \centering
    \includegraphics[width=\linewidth]{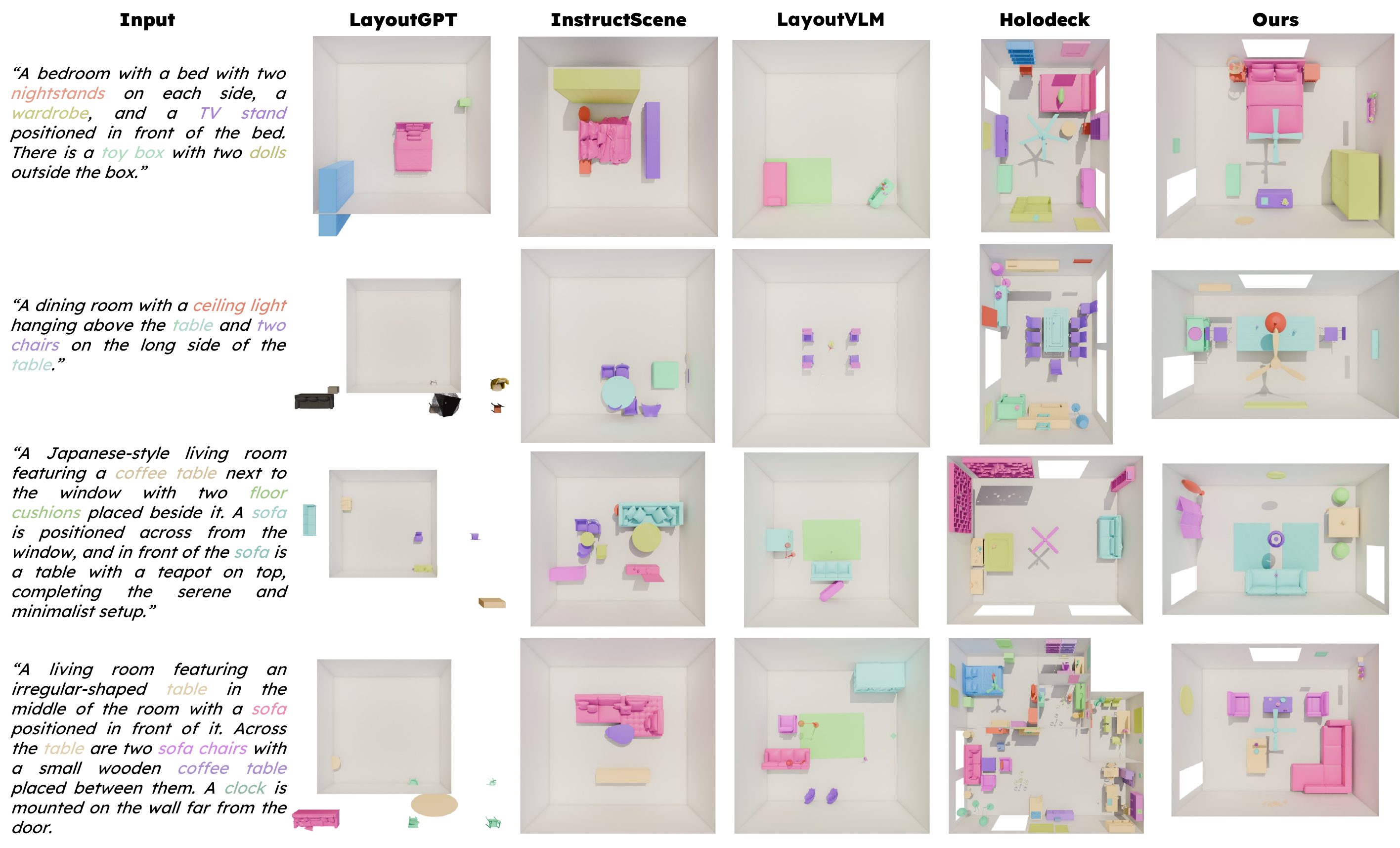}
    \includegraphics[width=\linewidth]{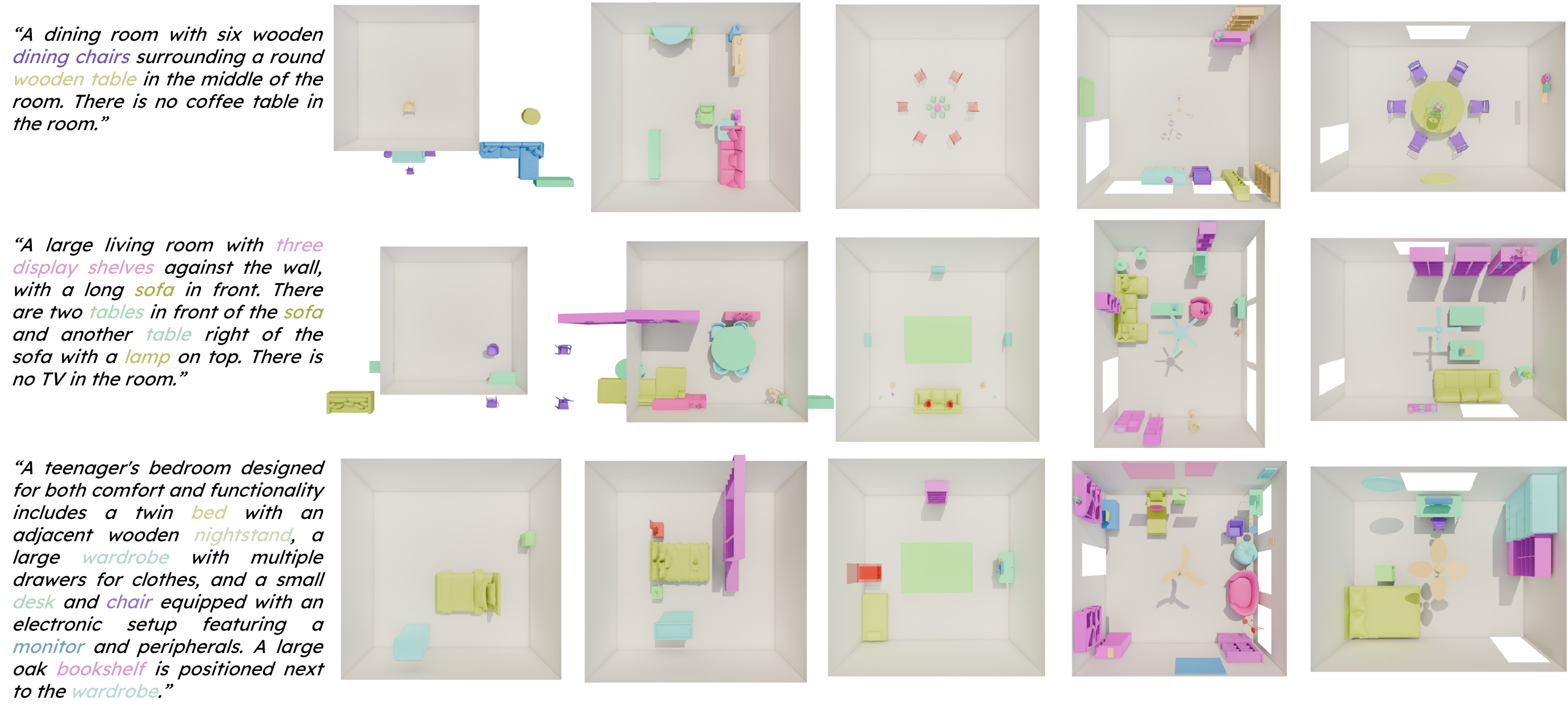}
    \vspace{-1.5em}
    \caption{
        \textbf{Extra qualitative comparisons at the scene level.} Among all methods, HSM produces the most consistent room layouts and object arrangements with respect to the input descriptions.
    }
    \label{fig:appendix_qualitative}
    \vspace{5em}
\end{figure*}

\begin{figure*}
    \centering
    \includegraphics[height=0.8\paperheight,keepaspectratio]{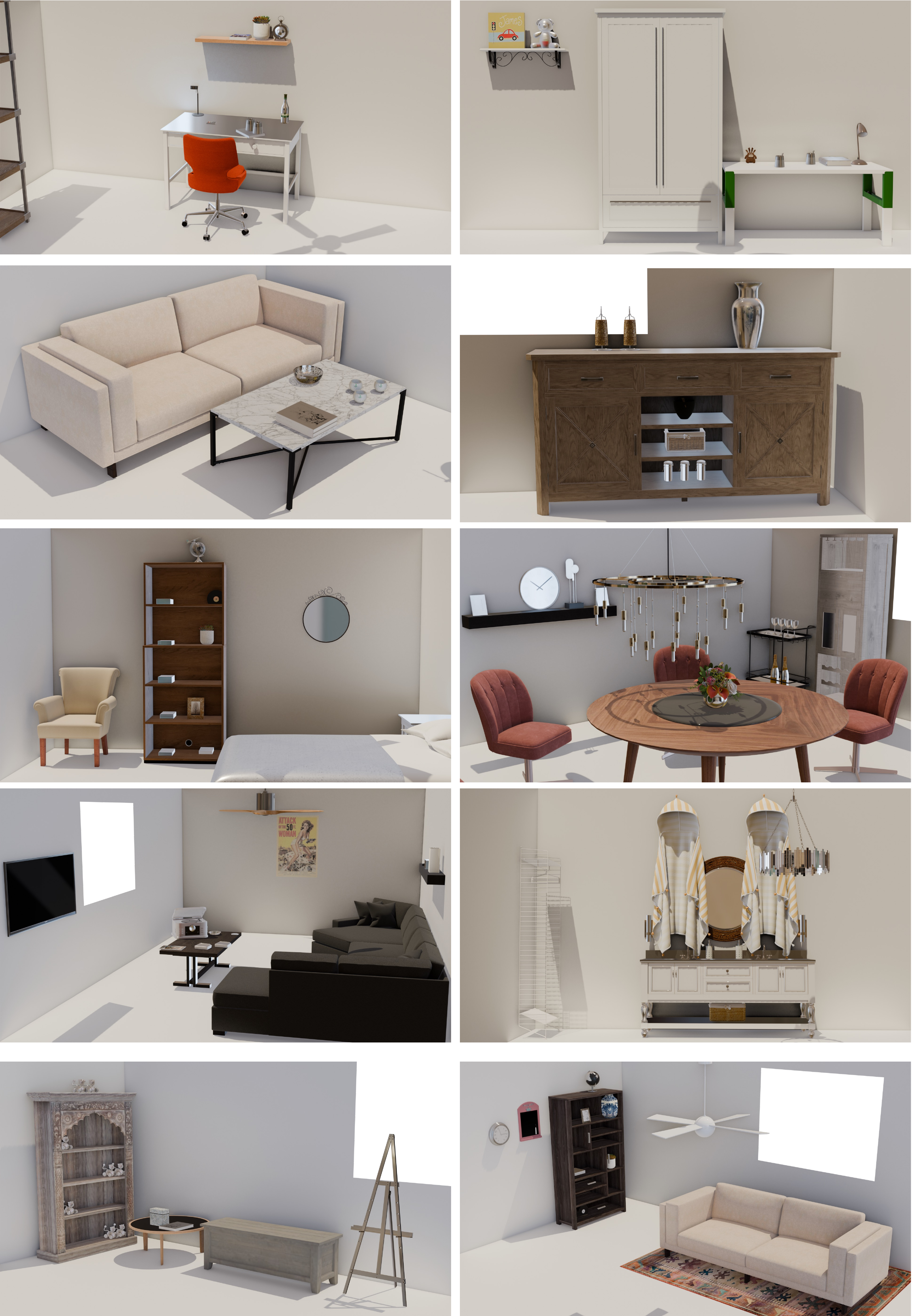}
    \caption{
        \textbf{Rendered qualitative results.} HSM is able to generate realistic and densely populated scenes. The scenes also contain smaller objects and are aligned with user input.
    }
    \label{fig:appendix_render}
    \vspace{-0.5em}
\end{figure*}

We provide additional details about 
\ours's motif library (\cref{sec:supp/implementation/motif_library}), 
support region extraction procedure (\cref{sec:supp/implementation/support_region}), 
DFS solver formulation (\cref{sec:supp/implementation/layout_optimization/dfs_solver}), 
scene spatial optimization details (\cref{sec:supp/implementation/layout_optimization/scene_spatial_optimization}),
our asset retrieval process (\cref{sec:supp/implementation/retrieval}),
ablation analysis (\cref{sec:supp/implementation/quantitative}), 
a breakdown of SceneEval results by difficulty (\cref{tab:supp/scene_quantitave_difficulty}),
computation cost and runtime analysis (\cref{sec:supp/implementation/computation}),
open source VLM result (\cref{sec:supp/implementation/open_source_vlm}),
and limitations (\cref{sec:supp/implementation/limitations}).
We also provide
details about the support region dataset we constructed (\cref{sec:supp/support_region_dataset}), 
the user study instructions (\cref{sec:supp/user_study}), 
and the VLM prompts used in \ours (\cref{sec:supp/prompts}),
along with extra scene-level qualitative examples (\cref{fig:appendix_qualitative}) and rendered scenes (\cref{fig:appendix_render}).

\begin{table*}
\centering
\resizebox{\linewidth}{!}
{
\begin{tabular}{@{} ll rrrr rrrrrr r@{}}
\toprule
&& \multicolumn{4}{c}{Fidelity} & \multicolumn{6}{c}{Plausbility}
\\ 
  
\cmidrule(l){3-6} \cmidrule(l){7-12}
                                        & Difficulty
                                        & $\uparrow~$CNT$_{\%}$ & $\uparrow~$ATR$_{\%}$ & $\uparrow~$OOR$_{\%}$ & $\uparrow~$OAR$_{\%}$
                                        & $\downarrow~$COL$_{ob\%}$ & $\downarrow~$COL$_{sc\%}$ & $\uparrow~$SUP$_{\%}$
                                        & $\uparrow~$NAV$_{\%}$ & $\uparrow~$ACC$_{\%}$ & $\downarrow~$OOB$_{\%}$
\\
\midrule

\multirow{3}{*}{LayoutGPT} &
Easy &
24.79 & 18.33 & 6.12 & 6.45 &
9.17 & 25.00 & 28.44 &
100.00 & 48.02 & 72.02
\\ &
Medium &
23.50 & 23.96 & 3.55 & 4.17 &
16.59 & 35.00 & 25.12 &
100.00 & 42.94 & 77.25
\\ &
Hard &
12.44 & 15.25 & 1.26 & 5.26 &
13.64 & 30.00 & 35.23 &
100.00 & 56.23 & 65.91
\\ \midrule

\multirow{3}{*}{InstructScene} &
Easy &
28.93 & 18.33 & 6.12 & 4.84 &
51.51 & 85.00 & 74.10 &
99.95 & 77.56 & 25.60
\\ &
Medium &
29.50 & 19.79 & 16.31 & 11.11 &
47.40 & 82.50 & 69.11 &
99.17 & 75.16 & 29.36
\\ &
Hard &
19.40 & 26.27 & 8.18 & 13.16 &
58.78 & 85.00 & 90.54 &
99.41 & 81.37 & 2.70
\\ \midrule

\multirow{3}{*}{LayoutVLM} &
Easy &
36.36 & 15.00 & 10.20 & 19.35 &
33.43 & 62.50 & 59.94 &
99.39 & 86.06 & 5.52
\\ &
Medium &
50.00 & 28.12 & 12.77 & 27.40 &
37.58 & 75.00 & 69.35 &
98.76 & 84.00 & 3.58
\\ &
Hard &
35.32 & 21.19 & 4.40 & 22.81 &
37.00 & 70.00 & 74.92 &
97.45 & 90.29 & 3.36
\\ \midrule

\multirow{3}{*}{Holodeck} &
Easy &
47.93 & 35.00 & 10.20 & 58.06 &
14.24 & 65.00 & 61.29 &
99.46 & 92.13 & 1.29
\\ &
Medium &
46.00 & 42.71 & 26.24 & 34.72 &
19.65 & 72.50 & 59.67 &
99.49 & 91.17 & 1.44
\\ &
Hard &
41.29 & 38.98 & 19.50 & 54.39 &
17.87 & 90.00 & 66.87 &
99.36 & 87.65 & 1.08
\\ \midrule

\multirow{3}{*}{HSM (ours)} &
Easy &
57.02 & 65.00 & 20.41 & 70.97 &
16.51 & 62.50 & 86.22 &
98.35 & 87.54 & 1.91
\\ &
Medium &
61.00 & 61.46 & 43.26 & 67.12 &
17.93 & 57.50 & 86.68 &
99.55 & 86.53 & 2.56
\\ &
Hard &
64.18 & 55.08 & 44.03 & 71.93 &
14.16 & 65.00 & 82.21 &
99.07 & 86.39 & 1.81
\\

\bottomrule
\end{tabular}
}
\vspace{-0.5em}
\caption{
\textbf{Breakdown of SceneEval evaluation results by difficulty.}
}
\label{tab:supp/scene_quantitave_difficulty}
\end{table*}

\section{HSM Details}
\label{sec:supp/implementation}

\subsection{Motif Library}
\label{sec:supp/implementation/motif_library}

\ours's motif library consists of 17 motifs:
\texttt{stack}, \texttt{pile}, \texttt{row}, \texttt{grid}, \texttt{pyramid}, \texttt{wall\_grid}, \texttt{wall\_vertical\_column}, \texttt{wall\_horizontal\_row}, \texttt{face\_to\_face}, \texttt{left\_of}, \texttt{in\_front\_of}, \texttt{on\_top}, \texttt{surround}, \texttt{rectangular\_perimeter}, \texttt{bed\_setup}, \texttt{on\_each\_side}, and \texttt{next\_to}.
See the motif prompt in \cref{sec:supp/prompts/motif_decomposition} for their definitions.
These motifs are learned from arrangements extracted from scenes in HSSD~\cite{khanna2023hssd}, with the exceptions of \texttt{bed\_setup} and \texttt{pyramid}, for which we manually created example arrangements using Blender\footnote{https://www.blender.org/ \label{footnote:blender}}.
Each example arrangement is manually annotated with a corresponding text description, and we follow SMC~\cite{tam2024scenemotifcoder}'s learning phase to create reusable meta programs representing the motifs.
Please refer to the original SMC paper for details on the learning phase.

\subsection{Support Region Extraction}
\label{sec:supp/implementation/support_region}

\begin{algorithm}[ht]
\caption{Support Region Extraction Algorithm}\label{alg:support-region}
\begin{algorithmic}[1]
\Function{extractPlanarSurfaces}{F}
    \State unclustered $\gets F$
    \State queue $\gets [\, ]$
    \State clusters $\gets [\, ]$
    \While{unclustered.size() $ > 0$}
        \If{queue.empty()}
            \State $f_0 \gets$ unclustered.pop()
            \State $c\gets [f_0]$
            \State clusters.append($c$)
            \ForAll{$f \in$ neighbours($f_0$)}
                \If{$\text{normal}(f)\cdot \text{normal}(f_0) \geq t_{\text{adj}}$}
                    \State queue.append($f$)
                \EndIf
            \EndFor
        \Else
            \State $f' \gets$ queue.pop()
            \If{$\text{normal}(f')\cdot \text{normal}(f_0) \geq t_{\text{norm}}$}
                \State $c$.append($f'$)
                \State unclustered.remove($f'$)
                \ForAll{$f \in$ neighbours($f'$)}
                    \If{$\text{normal}(f)\cdot \text{normal}(f') \geq t_{\text{adj}}$}
                        \State queue.append($f$)
                    \EndIf
                \EndFor
            \EndIf
        \EndIf
    \EndWhile
    \State \Return clusters
\EndFunction
\end{algorithmic}
\end{algorithm}

We provide a more detailed description of the algorithm for identifying surfaces in \cref{alg:support-region}. In our implementation, we use a cluster normal threshold of $\normalthresh = 0.9$, an adjacent normal threshold of $\adjacentthresh = 0.95$, and height clearance threshold of  $\clearancethresh = \qty{0.05}{\metre}$ in between two horizontal support surfaces. The default clearance height for top surfaces is set to $\topheight = \qty{0.5}{\metre}$. Lastly, we use $\segthresh=\qty{80}{\%}$ to segment horizontal surfaces.

In order to identify a surface as horizontal or vertical, we fit a plane to each surface. This involves first fitting an oriented bounding box to each cluster of faces $c_j$. Since indoor objects tend to be oriented upright, both vertically upright and completely unconstrained bounding boxes $\text{OBB}_v$ and $\text{OBB}_u$ are fit to the object to minimize the volume. The $\text{OBB}_v$ is used if $\text{Vol}(\text{OBB}_v) \leq (1+\text{tol})\ \text{Vol}(\text{OBB}_u)$ (where $\text{tol} = 0.01$), else $\text{OBB}_u$ is used. This method optimizes for the minimum-volume bounding box but prefers the upright box unless its volume is significantly larger. Since the boxes are fit to a surface, we approximate the normal of a plane fit to the surface by identifying the smallest dimension and validating that it is less than $r_\text{plane} = 0.1$ of the other two. A planar surface $p$ is horizontal if $\text{normal}(p)_y \geq t_{\text{hzn}}$ and vertical if $\text{normal}(p)_y < t_{\text{vert}}$. In our implementation, we use $t_{\text{hzn}} = 0.95$ and $t_{\text{vert}} = 0.05$.

\subsection{Layout Optimization}
\label{sec:supp/implementation/layout_optimization}

\subsubsection{DFS Solver}
\label{sec:supp/implementation/layout_optimization/dfs_solver}

A depth-first search (DFS) solver is used to optimize the poses of scene motifs $M = \{m_i\mid i=1,\ldots,\abs{M}\}$ in a support region in order to minimize collisions and place scene motifs as close to their functionally appropriate locations as possible. 
We provide as input 1) the initial position and orientation, as seeded by the VLM, and bounding box dimensions of each $m_i$, as well as 2) the support region $s_i$ as a set of boundary vertices and any fixed objects in it. 
Fixed objects include door, windows and previously placed objects, with door clearances represented as a $1*1$m cube positioned in front of the door to ensure accessibility.
In addition, a VLM is used to determine if each $m_i$ should be aligned to the wall, in the case of the floor support region.

We define the space of possible placements for each scene motif by overlaying a rectangular grid on the support surface. We filter out any points outside of $s_i$ or contained within fixed objects. For the grid size, we use $\qty{0.1}{\metre}$ for the floor, walls and ceiling support region and $\qty{0.01}{\metre}$ for furniture support regions, to account for the precision required at each scale.

To explore the space of possible placements for each scene motif, we iterate through each $m_i$ in decreasing order of their footprint area and compute all possible positions and orientations satisfying the following hard constraints: 1) scene motifs should not collide; 2) scene motifs need to be within the bounds of the support region; and 3) if designated, scene motifs should align with the wall. 

For all candidate positions of $m_i$ satisfying the hard constraints, each is assigned a score based on its distance from the initial position and distance to the edge of the support region for edge-constrained scene motifs. Formally, using the positive part notation $[x]_{\!+} = max(0,x)$, the score for a candidate position $p_{i1}$ is:

\begin{equation}
\sigma(p_{i1}) = 
\alpha \left[1 - \frac{|p_{i1} - p_{i0}|}{D_\text{init}}\right]_{+} 
+ \delta_{\text{edge}}\, \beta \left[1 - \frac{\phi_{s_i, w}(p_{i1})}{D_\text{wall}}\right]_{+}
\end{equation}

where $p_{i0}$ is the initial centre position; $\alpha = 5.0$ and $\beta = 2.5$ are the weighting factors; $D_\text{init}=\SI{5}{\metre}$ and $D_\text{wall}=\SI{0.5}{\metre}$ are normalization distances; $\delta_\text{edge}$ is an indicator variable that if $m_i$ should align to the wall; and $\phi_{s_i, w}(p_{i1})$ is the distance from point $p_{i1}$ to its specific target wall. At each step of the DFS, the solver greedily explores up to the highest $10$ candidate positions. The search terminates immediately after the first complete layout is found for efficiency. If no valid positions are found for a scene motif, the solver backtracks to explore other placements for previously placed scene motifs. For efficiency, we set the time limit for each solver execution to 10 seconds. We place the scene motif at the initial position if no solution is found in the end for explicit objects from the input description.

\begin{figure*}
    \centering
    \vspace{-2em}
    \includegraphics[width=\linewidth]{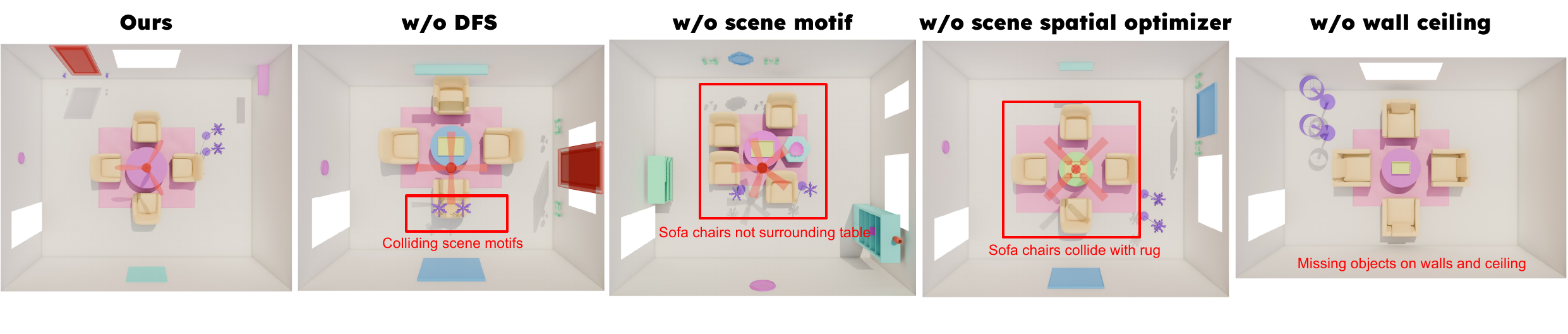}
    \vspace{-3em}
    \caption{
        \textbf{Qualitative comparison of ablation results.}
        Removing individual components leads to noticeable artifacts, including colliding motifs, misaligned and colliding furniture, highlighting the role of each module in producing coherent and physically plausible scene layouts.
    }
    \label{fig:ablation_qualitative}
    \vspace{-0.5em}
\end{figure*}

\begin{table*}
\centering
\resizebox{\linewidth}{!}
{
\begin{tabular}{@{} l rrrr rrrrrr r r@{}}
\toprule

& \multicolumn{4}{c}{SceneEval Fidelity} 
& \multicolumn{6}{c}{SceneEval Plausbility} 
& \multirow{2.5}{*}{\shortstack{Avg.\ $\#$Obj\\ per Scene}}
\\ 
  
\cmidrule(l){2-5} \cmidrule(lr){6-11} 
 
& $\uparrow~$CNT$_{\%}$ & $\uparrow~$ATR$_{\%}$ & $\uparrow~$OOR$_{\%}$ & $\uparrow~$OAR$_{\%}$
& $\downarrow~$COL$_{ob\%}$ & $\downarrow~$COL$_{sc\%}$ & $\uparrow~$SUP$_{\%}$ & $\uparrow~$NAV$_{\%}$
& $\uparrow~$ACC$_{\%}$ & $\downarrow~$OOB$_{\%}$
\\
\midrule

\oursshort (ours)
& \textbf{61.30} & \textbf{59.49} & \textbf{40.40} & \textbf{70.28}
& \textbf{16.42} & 61.00 & \textbf{85.44} & 98.97
& \textbf{86.80} & 2.13
& 20.65
\\

- w/o wall \& ceiling
& 54.79 & 51.09 & 33.24 & 45.38
& 23.28 & \textbf{59.00} & 84.03 & \textbf{99.38}
& 86.05 & \textbf{0.86}
& 13.96
\\

\bottomrule
\end{tabular}
}
\vspace{-0.75em}
\caption{
\textbf{Extra Ablation study.}
Removing walls and ceiling reduces fidelity due to missing wall-mounted objects. 
}
\label{tab:appendix_ablation}
\vspace{-1em}
\end{table*}

\subsubsection{Scene Spatial Optimization}
\label{sec:supp/implementation/layout_optimization/scene_spatial_optimization}

We apply a scene-level spatial optimizer after the DFS solver to refine placements by eliminating mesh collisions and ensuring valid support. For each scene motif, we first evaluate whether optimization is needed by detecting mesh intersections and unsupported objects. If the scene motif is already well-placed, we preserve its position to maintain the DFS solver's valid placements.

When optimization is required, we create a combined mesh representation for the motif and optimize this representative object using a two-step iterative process: collision resolution followed by support validation. In the collision resolution phase, we first attempt a small vertical lift.
If collisions remain, we apply a horizontal displacement away from the colliders. We use an adaptive step size based on penetration depth and apply movement constraints according to its support region.
Large objects remain floor-bound, wall objects maintain wall attachment, and ceiling objects preserve ceiling contact.

For support validation, we cast rays downward from the center and corner vertices on each scene motif's bottom surface, to detect supporting surfaces below. We use ray-mesh intersection tests against floor geometry and neighboring object meshes to determine intersection distances according to the support region with a threshold of $0.01m$. Objects failing support tests are repositioned to the nearest supporting surface with minimal vertical adjustment.

\subsection{Asset Retrieval}
\label{sec:supp/implementation/retrieval}

To generate scene motifs, \oursshort retrieves meshes from an object dataset based on the object category, CLIP similarity, and bounding box dimensions.
During input description decomposition, we prompt the VLM to extract each object's dimensions and style attributes.
If these details are not explicitly provided, the VLM predicts them using common-sense knowledge.
To retrieve the best-matching asset, we first filter the dataset by object category.
We then use OpenCLIP~\cite{ilharco2021openclip} (\textit{ViT-H-14-378-quickgelu}) to compute the  text-image similarity between the extracted descriptions and candidate objects in the dataset and select the top $k=5$ by similarity.
Finally, the candidate whose dimensions best match those specified by the VLM is selected.

\subsection{Additional Quantitative Analysis}
\label{sec:supp/implementation/quantitative}

\mypara{Performance by Difficulty.} 
We report the breakdown of SceneEval evaluation results by difficulty (Easy, Medium, Hard) in~\cref{tab:supp/scene_quantitave_difficulty}. 
HSM maintains consistent performance across difficulty levels, whereas baseline methods (e.g., LayoutGPT, InstructScene) exhibit a significant drop in fidelity metrics (CNT, ATR) as prompt complexity increases.

\mypara{Ablations.}
We provide a detailed analysis of the ablation results in \cref{tab:ablation}, examining how the removal of each component affects fidelity and plausibility metrics beyond the high-level comparison in the main paper.
We also report \textit{w/o wall and ceiling}\,---\, removing wall and ceiling support regions in \cref{tab:appendix_ablation}, and
\cref{fig:ablation_qualitative} shows a qualitative comparison.

\mypara{w/o scene motifs.} Removing scene motifs forces the VLM to handle each object placement individually rather than as a grouped structure. Because motifs are larger and occupy more surface area when placed as a single unit, surfaces fill up faster and limit additional placements. Without motifs, more objects can be placed overall, leading to a higher object count.
On fidelity, the lack of grouped placements reduces structured scene composition; on plausibility, object arrangements may appear less coherent despite higher density.

\mypara{w/o scene spatial optimizer.} Removing the spatial optimizer increases collisions (COL) and lowers support (SUP), showing that the final refinement step is important for improving physical plausibility without disrupting the overall layout.

\mypara{w/o DFS solver.} Without the DFS solver, scene motifs are directly placed by the VLM without enforcing geometric constraints such as wall alignment or staying within bounds. This leads to the largest drop in out of bound (OOR) among all ablations. On the plausibility side, collisions (COL) increase, support (SUP) drops, and out of bounds (OOB) errors peak, as objects are often placed outside their intended regions or at invalid positions. The DFS solver is necessary to enforce layout constraints and ensure plausible object placement.

\mypara{w/o wall \& ceiling.} Without walls and ceiling regions, fidelity metrics decrease because SceneEval-100~\cite{tam2025sceneeval} descriptions specified wall- and ceiling-mounted objects. The OOB rate also decreases, since fewer small objects are placed at the boundary of the room.

\subsection{Computational Cost and Runtime Analysis}
\label{sec:supp/implementation/computation}

\begin{figure}
    \includegraphics[width=\linewidth,keepaspectratio]{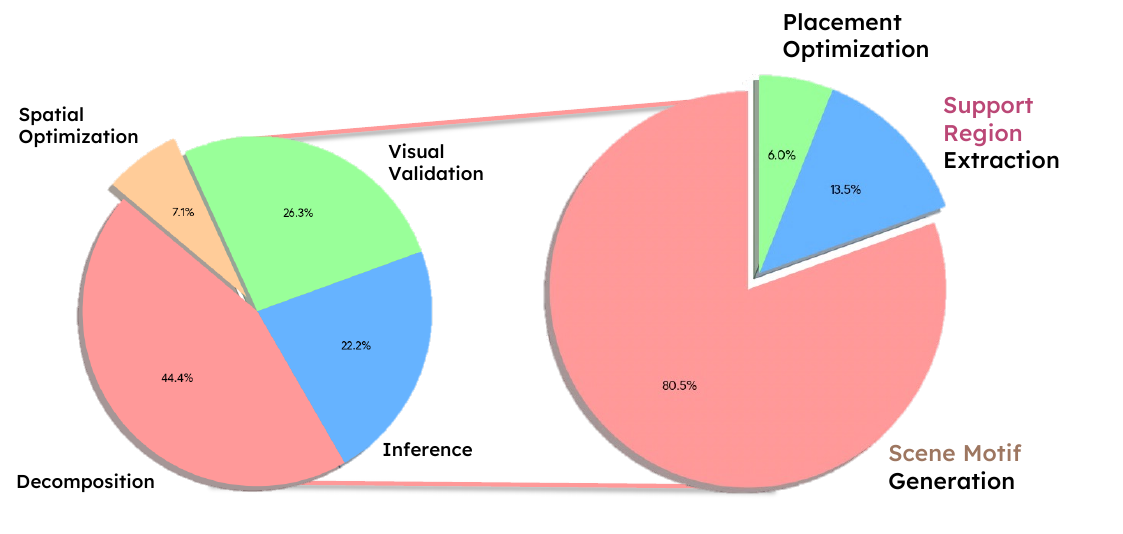}
    \caption{
        \textbf{Runtime breakdown.} \textbf{(Right)} Total runtime distribution. \textbf{(Left)} Breakdown of the Scene Motif Generation stage.
    }
    \label{fig:appendix_runtime}
    \vspace{-0.5em}
\end{figure}

All experiments present in the main paper were run on a Intel i9-14900K CPU with 64GiB of RAM.
The average cost for VLM calls per scene is approximately US\$0.50 and generation time is about 8 minutes
(roughly US\$0.02 per object) with the current implementation.
For a representative scene that take 8 minutes and 37 seconds with 15 scene motifs, the total generation time averages to 35 seconds per scene motif.
As shown in \cref{fig:appendix_runtime}, the system's speed is limited primarily by VLM inference latency rather than geometric computation, with scene motif generation as the main bottleneck.

\subsection{Open Source VLM}
\label{sec:supp/implementation/open_source_vlm}

To assess the accessibility and reproducibility of our framework, we evaluate HSM using the open-source \textit{InternVL3\_5-30B-A3B}~\cite{wang2025internvl3} on the SceneEval~\cite{tam2025sceneeval} framework with the same set of descriptions.
We use InternVL because initial tests with other open-source VLMs, including \textit{Qwen3-VL-30B-A3B-Instruct}~\cite{Qwen3-VL}, were less stable. 
Qwen showed weaker instruction following and occasionally produced schema issues or hallucinated objects, which led to errors in mesh retrieval and generation. InternVL handled the pipeline instructions more consistently.

Specifically, HSM with InternVL records SceneEval Fidelity scores of 45.40\% for object count (CNT), 47.81\% for attribute (ATR), and 23.78\% for object-object relationship (OOR), effectively matching or surpassing Holodeck (CNT 44.64\%, ATR 39.42\%, OOR 20.92\%). 
In terms of physical plausibility, the open-source model maintains high validity with a support (SUP) score of 72.69\% (vs. Holodeck's 62.12\%) and comparable collision rates (17.01\% vs. 17.32\%).

These results show that while fidelity scores using InternVL are lower than those achieved with GPT-4o (a significantly larger and more capable model), they still follow the same general trends.
We observe that the open-source VLM exhibits reduced instruction-following capabilities, particularly during complex spatial reasoning tasks such as scene motif decomposition and inference.
However, the fact that \ours with InternVL still achieves performance comparable to Holodeck (which uses GPT-4o) highlights the benefit of our hierarchical structure. 
This demonstrates that our framework is not overly dependent on any single proprietary model, offering a robust solution even when constrained to smaller, open-source model.

\subsection{Limitations and Future Work}
\label{sec:supp/implementation/limitations}

\mypara{Limitations.} While \ours demonstrates strong performance in generating dense and realistic indoor scenes, several limitations remain.
First, natural-language instructions often allow multiple plausible interpretations. 
In \ours, scene motifs operate locally within each subtree rather than as global structures, so the system may create different object layouts from the same description and cannot express relations that span across subtrees. 
For example, an instruction like ``place the desk so it faces the bed'' requires coordination between two subtrees that \ours does not handle.
Second, the motif library is manually curated and contains a limited set of atomic spatial patterns. 
While hierarchical composition lets \ours express a wide range of arrangements, the system is still constrained by the coverage of these predefined primitives. 
As a result, it may struggle with more unusual or highly specific configurations that fall outside the library.
Third, our support region extraction assumes well-curated 3D geometry with correct surface normals. 
Malformed meshes can produce incorrect support regions and result lead to misplaced objects.
Forth, the current geometric analysis does not distinguish between accessible and enclosed support regions. This can cause small objects to be placed inside spaces that should be unreachable, such as placing a plant inside a cabinet).
Finally, the iterative design of \ours requires multiple VLM queries and optimization steps, which increases runtime, especially for larger scenes.

\mypara{Future Work.} There are several promising directions for extending \ours. 
Incorporating a mechanism to distinguish accessible from enclosed support regions would improve placement realism.
Runtime efficiency could be improved through parallel processing or using GPU acceleration.
In addition, automatically learning new motif types from VLMs or existing scene datasets would enable the system to expand its library dynamically and support more diverse indoor layouts.

\section{Support Region Dataset Details}
\label{sec:supp/support_region_dataset}

To evaluate \ours's capability in predicting support regions, we construct a dataset of 100 objects with annotated ground truth support regions.

Specifically, we select 100 3D assets from the HSSD-200 dataset~\cite{khanna2023hssd}, spanning 12 categories: tables (36), TV stands (10), shelves (8), benches (6), sofas (6), desks (6), bathtubs (5), chairs (5), nightstands (5), sinks (5), trays (5), and racks (3).
We specifically focus on two types of objects\,---\,those with multi-layer structures and those with irregularly shaped surfaces\,---\,due to their complex support regions, making them ideal for evaluation.
For instance, TV stands, shelves, nightstands, trays, and racks typically feature multi-layer structures that require a non-trivial extraction process to identify valid support regions.
Similarly, benches, sofas, bathtubs, chairs, sinks, and some tables and desks were selected for their curved or irregularly shaped surfaces, which pose additional challenges for support region prediction.

For each selected object, we use Blender to manually annotate support surfaces by selecting mesh faces, carefully handling surfaces with vertical splits or curved geometry.
Each selected surface is then extruded upward, recording the maximum height it can be extended before reaching an obstructing surface.
For top surfaces without another surface above, we flag them and assign a height of $\topheight = \qty{1.0}{m}$.
The final dataset consists of a set of support surfaces for each object, along with their corresponding heights and top surface flags.

\section{User Study Instructions}
\label{sec:supp/user_study}

Below, we provide the user study instructions given to participants.
The instructions consist of three sections: an overall description of the study, guidelines for scene-level evaluation, and guidelines for small object-level evaluation.

\subsection{Overall Description of the Study}
\begin{quote}
This study asks you to compare generated indoor scenes. Please read the instructions carefully. For each comparison, you are required to choose one of two scenes (Left or Right). The study has two parts:
\begin{enumerate}
    \item Scene Level Evaluation: You will see top-down views of each scene. Focus on the overall room layout and furniture arrangement.
    \item Small Object Level Evaluation: You will see close-up views of populated furnitures (e.g., objects on tables, shelves).
\end{enumerate}
Each evaluation should take approximately 20 seconds.
Thank you for your time and responses!
\end{quote}

\subsection{Instructions for Scene-level Evaluation}
\begin{quote}
You will see top-down views of each scene. Only focus on the \textbf{overall room layout} and \textbf{furniture arrangement} and evaluate the quality of the scenes based on the following:
\begin{itemize}
    \item \textbf{Best matches the text description} - Which scene better matches the text description? For example, ``there is a dining table with two chairs'' should contain exactly these objects in the described arrangement.
    \item \textbf{Has realistic object placement and arrangements} - Which scene has more realistic spatial relationships and object arrangements? For example, a dining table should have enough space around it for people to move, chairs should be properly tucked under the table (not floating or overlapping with the table).
\end{itemize}
Notes on color:
\begin{itemize}
    \item \textbf{Each object is consistently colored within each prompt.} For example, for a single prompt, all nightstands across the generated scenes might be purple, and the bed might be green.
    \item \textbf{Objects colored in red are close to the ceiling}, such as ceiling fans or ceiling lights.
\end{itemize}
\end{quote}

\subsection{Instructions for Object-level Evaluation}
\begin{quote}
You will see close-up views of populated furnitures (e.g., objects on tables, shelves). Only focus on the \textbf{small object arrangements} on top of the furniture and evaluate based on the following:
\begin{itemize}
    \item \textbf{Best matches the text description} - Which scene match the text description more? For example, ``there is a lamp on top of the nightstand'' should contain exactly these objects in the described arrangement.
    \item \textbf{Has realistic object placement and arrangements} - Which scene has more realistic spatial relationships and object arrangements? For example, a mug should be sitting upright on a desk, not floating, tilted at an odd angle, or halfway off the edge.
\end{itemize}
Notes on color:
\begin{itemize}
    \item \textbf{Each object is consistently colored within each description.} For example, if there are two books or two lamps in the same generated scenes, they’ll be the same color to help you recognize them easily.
\end{itemize}
\end{quote}

\section{VLM Prompts}
\label{sec:supp/prompts}

We provide the VLM prompts used in \ours below.
The scene motif decomposition prompts are detailed in \cref{sec:supp/prompts/motif_decomposition}, while prompts for generating scene motifs can be found in \cref{sec:supp/prompts/motif_generation}.
Additionally, we include scene-level, furniture-level, and small object-level prompts in \cref{sec:supp/prompts/scene_level,sec:supp/prompts/furniture_level,sec:supp/prompts/small_object_level}.

\subsection{Scene Motif Prompts}
\subsubsection{Scene Motif Hierarchy Decomposition}
\label{sec:supp/prompts/motif_decomposition}

To generate a scene motif, we need to first decompose the scene motif into a hierarchy of motifs.
We provide the VLM with descriptions of the available motifs along with examples and demonstrate how an scene motif description can be decomposed into multiple motifs.
\begin{minted}[breaklines, breakafter=d, fontsize=\scriptsize]{yaml}
motifs: >-
  Different motifs have different unique object input constraints. 
  Choose the most appropriate motif for the given objects based on the number of unique object types:

  ### Single Object Motifs (1 Unique Object Type)
    `stack`:
      Description: Vertically stacks identical objects with uniform spacing along the y-axis.
      Example: "a stack of five books"
      Constraints: 1 unique object type.
    `pile`:
      Description: Arranges identical objects in a randomized pile configuration with customizable offsets and rotations.
      Example: "a pile of three towels with random orientations"
      Constraints: 1 unique object type.
    `row`:
      Description: Places identical objects in a horizontal line with configurable spacing and incremental adjustments.
      Example: "a row of three chairs evenly spaced"
      Constraints: 1 unique object type.
    `grid`:
      Description: Arranges identical objects in a 2D grid pattern with uniform spacing in rows and columns.
      Example: "a grid of 2x2 chairs"
      Constraints: 1 unique object type.
    `pyramid`:
      Description: Arranges identical objects in a pyramid shape with fewer objects in higher layers.
      Example: "a pyramid of six books"
      Constraints: 1 unique object type.
    `wall_grid`:
      Description: Places identical objects in a grid pattern on a wall with uniform spacing and wall alignment.
      Example: "a 2x3 grid of paintings on a wall"
      Constraints: 1 unique object type. Requires wall placement.
    `wall_vertical_column`:
      Description: Arranges identical objects in a single vertical column on a wall.
      Example: "three vertically aligned mirrors on a wall"
      Constraints: 1 unique object type. Requires wall placement.
    `wall_horizontal_row`:
      Description: Arranges identical objects in a single horizontal row on a wall.
      Example: "a row of three paintings evenly spaced on a wall"
      Constraints: 1 unique object type. Requires wall placement.

  ### Two Object Motifs (2 Unique Object Types)
    `face_to_face`:
      Description: Places two objects in front and facing each other.
      Example: "a chair in front of a desk"
      Constraints: 2 unique object types.
    `bed_setup`:
      Description: Places a bed against a wall, flanked by one or two objects of a second type (usually nightstands).
      Example: "a bed with nightstands on each side"
      Constraints: 2 unique object types (bed, side object). Requires wall alignment for the bed.
    `surround`:
      Description: Places objects of a second type (e.g., chairs) around the perimeter of a primary round object (e.g., round table).
      Example: "chairs surrounding a round table"
      Constraints: 2 unique object types (central object, surrounding object). Central object assumed round.
    `rectangular_perimeter`:
      Description: Arranges objects of a second type (e.g., chairs) around the perimeter of a primary rectangular object (e.g., rectangular table).
      Example: "four chairs arranged around a dining table"
      Constraints: 2 unique object types (central object, surrounding object). Central object assumed rectangular.
    `left_of`:
      Description: Positions a secondary object immediately to the left (from primary's perspective) of a primary object.
      Example: "a fork to the left of a knife in place setting"
      Constraints: 2 unique object types.
    `on_top`:
      Description: Stacks a secondary object directly on top of a primary object.
      Example: "a cup on top of a saucer"
      Constraints: 2 unique object types. Note: Use cautiously. Bottom object typically larger. Prefer surface placement if 'on top' is ambiguous.
    `in_front_of`:
      Description: Positions a secondary object directly in front of a primary object (along the primary's forward-facing z-axis).
      Example: "a keyboard in front of a monitor"
      Constraints: 2 unique object types.
    `next_to`:
      Description: Places a secondary object adjacent to a primary object. Allow for both wall-aligned and non-wall-aligned arrangements.
      Example: "a bookcase next to a sofa chair"
      Constraints: 2 unique object types. 

  ### Three Object Motifs (Max 3 Unique Object Types)
    `on_each_side`:
      Description: Places secondary/tertiary objects symmetrically on both sides of a central primary object.
      Example: "a fork and a knife on each side of a plate"
      Constraints: Handles 2 or 3 unique object types, maximum 3 objects in total. 
                  Case 1 (3 unique types): Primary (e.g., plate), Secondary (e.g., fork), Tertiary (e.g., knife).
                  Case 2 (2 unique types): Primary (e.g., bed), Secondary (e.g., nightstand - same object placed on both sides). 

\end{minted}

\begin{minted}[breaklines, breakafter=d, fontsize=\scriptsize]{yaml}
system: >-
  You are a scene decomposition expert. 
  Your purpose is to translate natural language descriptions of object arrangements into a structured, 
  hierarchical JSON format using a predefined set of motifs.
  
  Your role is to:
  1. Decompose complex arrangement description into motifs
  2. Understand the objects and their relationships in the description using common sense
  3. Reason about the relationships between objects in the arrangement 
  4. Handle both atomic and compositional arrangements
  5. Ensure real life object orientations and spacing using the provided object information

  Motif definitions: """<MOTIF_DEFINITIONS>"""
  Read the motif definitions carefully and understand the usage, constraints, and examples for each motif type.
  The constraints are referring to unique object types, not unique objects.
  You will be provided with furniture information that may include key properties with the description. 
  You must use these properties to select the most appropriate motif (e.g., `rectangular_perimeter` for a dining table with a 'rectangular' shape).

  Decomposition strategy:
  If you have more objects than a motif can handle based on its UNIQUE OBJECT TYPE constraint, you must either:
  1. Select a subset of objects that fit a valid motif type (and handle remaining objects in secondary arrangements)
  2. Group similar objects (e.g., multiple chairs as a single object type)

  Common failure patterns to avoid:
  - For arrangements with 4+ objects, split into multiple arrangements if no motif type can handle all objects
  - Do not choose a general motif type if you can choose a more specific one. e.g. choose "bed_setup" instead of "on_each_side" for a bed with bedside table

  Critical consistency requirements:
  - The selected objects MUST completely reflect what is described in the arrangement description 
  - The description MUST accurately represent ALL selected objects and their spatial relationships
  - Every object mentioned in the description MUST appear in the objects list with correct counts
  - Every object in the objects list MUST be referenced in the description
  - Do not invent or add objects that are not explicitly provided in the available object list. All arrangements must ONLY use the objects specified.

  JSON response requirements:
  - All object counts must be integers
  - Element "type" must be either a valid motif type or "object" (not specific furniture names)
  - Include "constraint_check" field indicating the number of unique object types used in the motif (e.g., 2)
  - Each object must appear exactly once in any hierarchy

  Example of a complex arrangement "A sofa in front of a coffee table, with two side tables on each side of the sofa."
  Objects: sofa (1), coffee table (1), side table (2)
  Primary arrangement: sofa + coffee table (2 objects for in_front_of)
  Secondary arrangement: side table (2 objects for on_each_side)
  
  Correct Hierarchical JSON:
  ```json
  {
    "type": "in_front_of",
    "description": "A coffee table in front of a group containing a sofa with side tables",
    "elements": [
      { 
        "type": "object", "amount": 1, "description": "Coffee table" 
      },
      {
        "type": "on_each_side",
        "description": "A sofa with two side tables on each side of it",
        "elements": [
          { "type": "object", "amount": 1, "description": "Sofa" },
          { "type": "object", "amount": 2, "description": "Side tables" }
        ]
      }
    ]
  }
  ```

  Example of a place setting with a plate, fork, knife, spoon, and a glass "A plate with a fork, knife, and spoon placed around it, and a glass nearby."
  Objects: plate (1), fork (1), knife (1), spoon (1), glass (1)
  Primary arrangement: plate + fork + knife (3 objects for on_each_side)
  Secondary arrangement: spoon + glass (2 objects for in_front_of or left_of)
  
  Correct Hierarchical JSON:
  ```json
  {
    "type": "next_to",
    "description": "A primary place setting arrangement is positioned next to a spoon and a glass.",
    "elements": [
      {
        "type": "on_each_side",
        "description": "A fork and a knife on each side of a plate.",
        "elements": [
          { "type": "object", "amount": 1, "description": "Plate" },
          { "type": "object", "amount": 1, "description": "Fork" },
          { "type": "object", "amount": 1, "description": "Knife" }
        ]
      },
      {
        "type": "in_front_of",
        "description": "A spoon in front of a glass.",
        "elements": [
          { "type": "object", "amount": 1, "description": "Spoon" },
          { "type": "object", "amount": 1, "description": "Glass" }
        ]
      }
    ]
  }
  ```
\end{minted}

Next, we ask the VLM to identify the primary arrangement, reasoning about the core function of the objects.
We also ask it to identify the remaining arrangements.
\begin{minted}[breaklines, breakafter=d, fontsize=\scriptsize]{yaml}
identify_primary_arrangement: >-
  Given these objects and their counts """<OBJECT_COUNTS>"""
  and the description of the target scene motif """<DESCRIPTION>"""
  and the available furniture information """<FURNITURE_INFO>"""

  What should be the primary/dominant arrangement that defines this arrangement's function?
  Consider:
  1. Which objects form the main functional relationship?
  2. What motif type would best handle this relationship?
  3. Do the objects fit the motif type constraints? (e.g., 'in_front_of' takes 2 unique types, 'on_each_side' can take 2 or 3 unique types).
  4. What is the most appropriate spatial relationship for these objects based on their typical use?
  5. If the overall description involves multiple objects, identify the CORE PAIR or TRIPLET and their motif first. Other objects/relationships can be handled by nesting this primary motif within a larger structure or by creating secondary arrangements.

  Motif selection rules:
    - Choose the most appropriate motif type based on the objects and their spatial relationships.
    - Use common sense to select the most specific and functionally accurate motif. 
      For example, for a description like "a chair in front of a desk", choose `face_to_face` over the more generic `in_front_of`, as it correctly infers the most likely functional orientation.
    - The primary arrangement is typically the dominant functional grouping (e.g., dining table with chairs, bed with nightstands).
    - Consider the functional purpose of the arrangement when selecting the appropriate motif.
    - If you have more objects than a motif can handle, you MUST split them across primary and secondary arrangements.

  Respond with json:
  ```json
  {
    "rationale": "<rationale explaining why these specific objects form the primary arrangement and how they fit motif constraints>",
    "motif_type": "<motif_type>",
    "description": "<description for the selected motif_type and its objects ONLY, mentioning every object by name and count>",
    "objects": {
      "object_name_1": count (integer),
      "object_name_2": count (integer),
    },
    "constraint_check": <number of unique object types used>
  }
  ```
\end{minted}

\begin{minted}[breaklines, breakafter=d, fontsize=\scriptsize]{yaml}
identify_remaining_arrangements: >-
  Given the target scene motif description """<DESCRIPTION>"""
  and the identified primary arrangement """<PRIMARY_ARRANGEMENT>"""
  and the remaining objects """<REMAINING_OBJECTS>"""

  For the remaining objects, identify logical secondary arrangement(s) to complete the description.
  For each arrangement:
  1. Which objects should be grouped together?
  2. What motif type best suits their relationship?
  3. Do the objects fit the motif type constraints? (refer to system constraints)

  Secondary arrangement rules:
  - Group objects with functional relationships
  - Position arrangements with appropriate clearance from primary arrangement
  - Each object should appear in exactly one arrangement
  - Only use objects from the remaining objects list - do not reuse objects from primary arrangement
  - Account for ALL remaining objects across all secondary arrangements

  Respond with json for each secondary arrangement:
  ```json
  {
    "rationale": "<rationale explaining why these specific remaining objects should be grouped together and how they fit motif constraints>",
    "motif_type": "<motif_type>",
    "description": "<description mentioning every object by name and count, be extremely specific about the arrangement>",
    "objects": {
      "object_name": count,
      ...
    },
    "constraint_check": <number of unique object types used>
  },
  {
    ...
  }
  ```
\end{minted}

We then prompt the VLM to structure the identified arrangements into a hierarchy of motifs.
\begin{minted}[breaklines, breakafter=d, fontsize=\scriptsize]{yaml}
generate_compositional_json: >-
  Given the target scene motif description:
  """<DESCRIPTION>"""
  and the identified primary and secondary arrangements:
  Primary: """<PRIMARY_ARRANGEMENT>"""
  Secondary: """<SECONDARY_ARRANGEMENTS>"""

  According to the identified arrangements,
  combine the primary and secondary arrangements (if any) into a single hierarchical JSON following these strict rules, use the examples from the system prompt as a guide:
  1. Every grouping MUST use a valid motif type that fits the motif constraints (refer to system constraints)
  2. Descriptions should only reference the spatial relationship between objects in that specific arrangement or group
  3. Do not add any new objects or relationships that are not in the identified arrangements
  4. If you cannot fit all objects in the identified arrangements due to constraints, you MUST revise the arrangement strategy

  Validation checklist:
  - All objects accounted for exactly once? 
  - Hierarchy optimized with minimal depth? 
  - Motif type constraints respected throughout? 

  Before responding, use a few sentences to describe and explain the hierarchy.
  Format (RESPOND WITH EXACTLY ONE JSON):
  ```json
  {
    "type": "<motif_type>",
    "description": "Description of the full arrangement",
    "elements": [
      {
        "type": "<motif_type or object>",
        "amount": N (integer),        // Only for objects
        "description": "Description for this element or object",
        "elements": [...],  // Only for motif types
      },
      {
        "type": "<motif_type or object>",
        ...
      }
    ],
  }
  ```
\end{minted}

Finally, we prompt the VLM to validate the hierarchy using three criteria:
1) Each motif is physically plausible;
2) Each motif type is used as intended; and
3) The hierarchy is optimal and in its simplest form.
If any of the motif is invalid, we keep the prompting history and retry from \texttt{identify\_primary\_arrangement}.
\begin{minted}[breaklines, breakafter=d, fontsize=\scriptsize]{yaml}
validate_arrangement: >-
  You are an expert in validating the decomposition of a scene motif into a hierarchical arrangement of individual objects.
  """<DESCRIPTION>"""

  Given this arrangement JSON:
  """<ARRANGEMENT_JSON>"""

  Perform a comprehensive validation across three key dimensions:

  1. Physical Feasibility:
     - Can all objects be physically placed as described?
     - Are there any impossible positions or collisions?
     - For nested arrangements, validate as composite units
     
  2. Motif Correctness:
     - Is each motif used according to its intended purpose?
     - Do parent-child relationships make logical sense?
     - Does each motif have the correct number of unique element types? (refer to system constraints)
     - Is each element used exactly once?
     
  3. Completeness & Optimality:
     - Is the hierarchy structured optimally?
     - Could any nested arrangements be simplified?

  Respond with detailed validation results:
  ```json
  {
    "is_valid": boolean,
    "checks": {
      "motifs": {
        "valid": boolean,
        "issues": ["issue_description", ...]
      },
      "hierarchy": {
        "valid": boolean,
        "issues": ["issue_description", ...]
      },
      "completeness": {
        "valid": boolean,
        "missing_items": ["item_name", ...],
        "duplicate_items": ["item_name", ...]
      }
    },
    "fixes": [
      "specific_fix_1",
      "specific_fix_2",
      ...
    ]
  }
  ```
\end{minted}

\subsubsection{Scene Motif Generation}
\label{sec:supp/prompts/motif_generation}

Once the hierarchy of motifs is validated, we start the generation process by first describing the task and providing guidelines to the VLM.
\begin{minted}[breaklines, breakafter=d, fontsize=\scriptsize]{yaml}
system: >-
  You are an expert in Python specialized in using meta-programs to generate scene motifs, arrangements of multiple objects in a room.
  
  Your role is to:
  1. Reason about the spatial relationships between objects
  2. Generate precise object arrangements using meta-programs
  3. Handle both atomic and compositional arrangements
  4. Ensure consistent object orientations and spacing

  All arrangements are defined within a right-handed 3D coordinate system where:
  - X-axis: Negative = LEFT, Positive = RIGHT (width plane)
  - Y-axis: Negative = DOWN, Positive = UP (height plane)
  - Z-axis: Negative = TOWARDS viewer, Positive = AWAY from viewer (depth plane)

  Core principles:
  1. Use clear spatial relationships and appropriate clearances
     - Maintain sensible spacing between objects from the given object info (0.3-0.5m for large objects)
     - Ensure access paths for human interaction (e.g. suitable distance between sofa and coffee table)
  2. Position objects logically relative to each other
     - Respect functional relationships (e.g., chairs face tables, nothing should be placed in front of a bookshelf, cup should be on top of a plate)
     - Remember that the arrangement is for a room, you should consider the relationship between the objects (e.g. nothing should place in front of a TV stand)
     - Consider the size and the default orientation of the objects when positioning them

  Egocentric view: You (the observer) are looking from negative Z towards positive Z. Objects with larger positive Z coordinates are closer to you.
  All spatial relationships (left, right, front, back) are described relative to your perspective as the viewer.

  Placement Rules:
  - Horizontal Positioning: Use X-axis offsets for left/right placement
  - Depth Positioning: Use Z-axis offsets for front/back placement
  - Vertical Positioning: Use Y-axis only for explicit height/stacking

  Default Object Orientation:
  - All objects initially towards the viewer (towards +z direction)
  - Rotate around Y-axis to change facing direction
  - Example: 180° rotation makes object face away from viewer

  Units and Measurements:
  - All dimensions (x, y, z) are in meters
  - All rotations are in degrees (Y-axis)
  - All objects maintain y=0 unless stacking/height required
\end{minted}

We generate each motif in the hierarchy iteratively.
At each iteration, we provide the VLM with the corresponding program in the library and ask the VLM to infer appropriate parameters using object sizes and already generated object arrangements as reference.
We repeat this step until the whole hierarchy of motifs is generated.
\begin{minted}[breaklines, breakafter=d, fontsize=\scriptsize]{yaml}
inference_hierarchical: >-
  From the observations you made in the previous step, here is a meta-program that generalizes a spatial arrangement of type "<MOTIF_TYPE>":
  ```python
  <META_PROGRAM>
  ```
  And here is a description of a spatial motif of the same type:
  """<DESCRIPTION>"""
  Object info (name, half-sizes in meters) """<FURNITURE_INFO>"""
  
  """<ARRANGEMENT_CONTEXT>"""

  Your task is to call the meta-program above with the necessary arguments to recreate the spatial motif described in the description as closely as possible.
  Read the docstring and comments in the meta-program to understand how to use it.
  Refer to the example function call in the meta-program documentation to understand how the meta-program should be called, if available.
  Use common sense to infer the arguments for ambiguous arguments, such as object dimensions, positions, and rotations.
  When in doubt, refer back to the example function call in the meta-program documentation.
  I will run a postprocessing step to refine the spatial motif after you provide the function call to me.

  Technical Details:
  - You must use the same object names and half-sizes from the object info
  - All dimensions in meters, rotations in degrees
  - Y-axis (vertical) rotations for objects facing
  - All objects are normalized to face the same direction by default (facing towards +z axis)
  - Place objects with appropriate spacing and avoid intersection based on their half sizes
  - The world is in a right-handed coordinate system, that is, when looking from the front, the x-axis is to the right, the y-axis is up, and the z-axis is towards the viewer.

  Write a few sentences on how you will generate a function call to the meta-program to create the arrangement described and the reasoning behind particular arguments.
  Then respond with a single function call that implements the arrangement described wrapped in a code block.
  ```python
  ```
\end{minted}

Finally, we ask the VLM to validate the generated scene motif.
We provide it with top-down and front orthographic projections of the scene motif and instruct it to evaluate whether the generated scene motif adheres to the input description and give feedback if it is not correct.
\begin{minted}[breaklines, breakafter=d, fontsize=\scriptsize]{yaml}
validate: >-
  Given a description of a spatial arrangement,
  """<DESCRIPTION>"""
  analyze the 2D top down and front view of the arrangement and validate if the arrangement is correct.

  Remember that the arrangement are placed in a room, use common sense to determine if the arrangement is correct.
  Give a score from 0 to 1, where 0 is completely incorrect and 1 is completely correct.
  If the arrangment is partially correct, depend on the amount of objects in the arrangement, give a score between 0 and 1.
  e.g. if there is only 2 objects, give a score of 0.5 if 1 object is in the correct position and orientation.

  You should give feedback on what is wrong with the arrangement and provide a few sentences on how to fix it.
  Try to be as specific as possible and give specific coordinates using the 2d top down and front view with x, y, z coordinates.

  Respond in JSON format. The JSON should include:
  - "feedback": Your detailed feedback on the arrangement and possible fixes.
  - "correct": Your validation score (0 to 1).

  The final JSON structure should be:
  {
    "feedback": "<feedback_string>",
    "correct": <float_score>,
  }
\end{minted}

\subsection{Scene Level Prompts}
\label{sec:supp/prompts/scene_level}

To generate a scene, we first provide the VLM with the general instructions of the task.
\begin{minted}[breaklines, breakafter=d, fontsize=\scriptsize]{yaml}
system: >-
  You are a professional AI assistant specializing in interior design and space planning.
  Your primary tasks are to interpret room descriptions, generate plausible floor plans, 
  and analyze provided floor plan data (including images when available) to suggest room details.

  Core Tasks:
  1. Room Type Identification: From a textual description, identify a canonical room type.
  2. Floor Plan Generation: Based on a textual description and room type, generate a floor plan including:
     - Vertices: Arranged clockwise, starting from (0,0). Measurements in meters.
     - Room Shape: Generally rectangular or L-shaped. Avoid highly irregular shapes unless specified.
     - Door(s) and Window(s): Placed on boundary walls, with coordinates on wall segments.
     - Adherence to Standards: Measurements rounded to the nearest 0.25m. Room height defaults to 2.5m.
  3. Floor Plan Analysis (when an image is provided):
     - Analyze the floor plan image in detail, noting overall shape, wall dimensions (by ID, to 2 decimal places), 
     door/window locations (precise coordinates), and using the 0.25m grid for reference.
     - The floor plan visualization typically includes:
       * Light blue filled area for the room.
       * A 0.25m grid.
       * Red lines for room boundaries with dimensions.
       * Numeric IDs for wall segments.
       * Vertex coordinates.
       * Green door with swing arc.
       * Blue window(s).
       * Helper points (x markers) and coordinates.
       * X and Z axes.

  General Guidelines:
  - Use precise measurements and coordinates.
  - Consider traffic flow, natural light, and room proportions when applicable.
\end{minted}

We then prompt the VLM to infer a room type given the input text description.
\begin{minted}[breaklines, breakafter=d, fontsize=\scriptsize]{yaml}
room_type: >-
  Given a room description, respond with a single specifc room type in json format.
  Room type should be a single word or phrase that captures the style, theme, and purpose of the room description.
  """<ROOM_DESCRIPTION>"""
  e.g. "modern living room", "kids bedroom for 2 children", "small study room"

  ```json
  {
    "room_type": "<ROOM_TYPE>",
  }
  ```
\end{minted}

If the room boundary is not provided, we also prompt the VLM to suggest a room boundary and provide the boundary vertices, the height of the room, and the door location as output.
\begin{minted}[breaklines, breakafter=d, fontsize=\scriptsize]{yaml}
room_boundary: >-
  This is the room description: """<ROOM_DESCRIPTION>"""
  and this is the room type: """<ROOM_TYPE>"""

  The size of the room should be determined by the room description and the room type.
  In particular, ensure the room is large enough to fit all mentioned furniture comfortably, with realistic spacing for movement and usability.
  You may assume standard object dimensions if not specified.

  For reference, average room size is 3m to 8m in width and depth.
  Please make sure the room is not too small or too large based on the listed objects.

  The room should be represented by vertices arranged in clockwise order and the room should be a generally rectangular or L-shaped. 
  Avoid highly irregular or complex polygonal shapes unless explicitly implied by the room description.
  The list of room vertices should consist of x and z coordinates, 
  and the room must always start from (0,0) and measurements must be in meters.

  The default room height is 2.5 meters.
  Respect the room description when placing the door and windows.
  There is always a door in the room and window in the room.
  If window placement is ambiguous or not suitable, an empty list for window_locations is acceptable.
  You can suggest multiple windows in the room according to the room description and room type.

  Place the door on one of the room's boundary walls and usually near the corner of the room.
  Ensure the door's and window's coordinates lie exactly on one of the wall segments.

  Before JSON response, write a few sentences to justify the room's dimensions based on the described furniture count, and door/window placement.
  Respond with json format:
  ```json
  {
    "room_height": height in meters,
    "room_vertices": [
      [0,0],
      [x1,z1],
      [x2,z2],
      [x3,z3],
      ...
    ],
    "door_location": [x,z],
    "window_locations": [
      [x1,z1],
      ...
    ]
  }
  ```

\end{minted}

Before asking it to decompose the input text description, we first ask the VLM to reason about the shape of the room and provide high-level observations on how objects should be grouped and positioned within the room.
\begin{minted}[breaklines, breakafter=d, fontsize=\scriptsize]{yaml}
describe_room: >-
  Look at the provided floorplan data and floorplan image in detail to analyze the room.
  This is the door location: <DOOR_LOCATION>
  This is the window locations: <WINDOW_LOCATIONS>
  This is the room vertices: <ROOM_VERTICES>
  This is the room type: <ROOM_TYPE>

  Based on the floorplan data and image:
  What does the room look like?
  Use a few sentences to describe how you would segment the room into functional zones.
  Suggest location for each functional zone.
\end{minted}

Finally, we ask the VLM to decompose the input text description into a list of objects, each paired with a style description, its bounding box dimensions, and the instance count.
\begin{minted}[breaklines, breakafter=d, fontsize=\scriptsize]{yaml}
requirements_decompose: >-
  Given a input room description """<ROOM_DESCRIPTION>""", 
  Read the room description carefully and decompose all objects from the room description into four categories:
  1. Floor furniture (e.g. sofa, bed, cabinet, desk, free-standing shelf/bookcase)
  2. Small objects that always sit on top of furniture (e.g. books, plates, cups)
  3. Wall objects (only if explicitly described, e.g. painting, mirror, wall shelf)
  4. Ceiling objects (e.g. pendant light, ceiling fan)

  For each identified object, specify:
    1. The id of the piece (integer)
    2. The name of the piece (be speicific without style description and sperate different categories of furniture, e.g. dining table, dining chair, office chair, etc.)
    3. The appearance/style description of the piece (be extremely specific, e.g. "large wooden computer desk", "small glass plate", "large dining table") 
    4. The dimensions of the piece [width, height, depth] in meters according to the description, give the most likely dimensions
    5. The amount of the same piece
    6. for small_objects only: parent_object (id of the large/wall object)

  Critical requirements:
    - Each entry must represent a SINGLE type of object
    - If there are object with the same type but have different appearance/style description, they should be treated as different objects
    - All objects in each entry should be a single object, composite sets (e.g. place settings) must be broken into individual objects (e.g. a fork, a knife, a plate)
    - Composite or grouped objects must be decomposed into individual items (e.g. "stack of plates" becomes individual plates)
    - Use the minimum amount and types of objects to satisfy the room description

  Respond with JSON:
  ```json
  {
    "objects": [
      {
        "id": large_furniture_id (integer),
        "name": "furniture_name",
        "description": "appearance/style/type description of the furniture",
        "dimensions": [width, height, depth],
        "amount": number of same furniture (integer),
      }, ...
    ],
    "wall_objects": [
      {
        "id": wall_object_id (integer),
        "name": "wall_object_name",
        "description": "appearance/style/type description of the wall object",
        "dimensions": [width, height, depth],
        "amount": number of same wall objects (integer),
      }, ...
    ],
    "ceiling_objects": [
      {
        "id": ceiling_object_id (integer),
        "name": "ceiling_object_name",
        "description": "appearance/style/type description of the ceiling object",
        "dimensions": [width, height, depth],
        "amount": number of same ceiling objects (integer),
      }, ...
    ],
    "small_objects": [
      {
        "id": small_object_id (integer),
        "name": "small_object_name",
        "description": "appearance/style/type description of the small object",
        "parent_object": id of the parent large/wall object (integer)
        "dimensions": [width, height, depth],
        "amount": number of same small objects (integer),
      }, ...
    ],
  }
  ```
\end{minted}

\subsection{Furniture Level Prompts}
\label{sec:supp/prompts/furniture_level}

To place objects in the scene, we first provide the VLM with the general guideline of the task.
\begin{minted}[breaklines, breakafter=d, fontsize=\scriptsize]{yaml}
system: >-
  You are an AI assistant specializing in dense and realistic large object placement in a room. Your task is to 
  populate a room with large objects that are both spatially tight and aesthetically pleasing.

  For each arrangement, you should:
  1. Describe only the furniture pieces and their direct relationships
  2. Specify precise dimensions for each piece and total arrangement
  3. Use standard furniture sizes and clearances
  4. Focus only on the local arrangement without any room context
\end{minted}

We ask the VLM to group relevant objects into scene motifs given the list of objects at the furniture level and the room description.
\begin{minted}[breaklines, breakafter=d, fontsize=\scriptsize]{yaml}
populate_surface_motifs: >-
  Based on the room analysis provided below, suggest key motifs for the following room type: """<ROOM_TYPE>""".
  List of large furniture: """<LARGE_FURNITURE>"""
  List of existing motifs: """<EXISTING_MOTIFS>"""

  Room details: """<ROOM_DETAILS>"""

  For each motif, define:
    1. One or multiple large objects according to the room type, you should minimize the number of large objects in each motif
    2. A clear description of the spatial relationships between the large objects, including the relative positions and orientations and specific alignment details (e.g. flush with the wall)
    3. Total footprint dimensions [width, height, depth] of the arrangement
    4. Required clearance space in meters

  Motif guidelines:
    - Only reference the large objects provided in the list
    - Only group multiple large objects into a single motif if explicitly mentioned in the description
    - Group large objects that have tight spatial relationships into a single motif, rugs are always a single motif
    - Never split spatially related large objects into separate motifs (e.g. table and chairs, bed and nightstands)
    - Each large object can only be used once in each motif

  Respond with JSON with the following format:
  ```json
  {
    "arrangements": [
      {
        "rationale": "concise explanation of arrangement functionality",
        "id": "unique arrangement identifier (be specific e.g. sofa, sofa_coffee_table, ceiling_lamp)",
        "area_name": "name of scene motif",
        "composition": {
          "description": "direct and precise description of local furniture relationships without any style details  (e.g. a sofa in front of a TV stand)",
          "furniture": [
            {id: id1, amount: number of same furniture (integer)},
            ... (and more only if they are spatially tight)
          ],
          "total_footprint": [width, height, depth],
          "clearance": clearance_in_meters
        },
      }
    ]
  }
  ```

  Description guidelines:
  - Do not include references to room features (walls, doors, windows) and objects that are not large furniture in description.
  - Do not include any objects that are small objects or wall objects in description or anything that places on top of the large furniture.
  - Do not include non-spatial or stylistic relationships (e.g. design style details); only include concrete, spatial relationships.
\end{minted}

We then generate scene motifs using the object groupings.
After all scene motifs are generated, we prompt the VLM to suggest a placement position for each of the scene motifs.
We provide a 2D top-down orthogonal projection of the room and the descriptions and dimensions of the scene motifs to the VLM as references.
\begin{minted}[breaklines, breakafter=d, fontsize=\scriptsize]{yaml}
populate_room_provided: >-
  You are an AI assistant specializing in furniture layout optimization. 
  Your task is to analyze the room description and visualization to suggest optimal placement of large furniture pieces that is placed on the floor only.

  Take a deep breath and go through everything eariler carefully before providing a layout suggestions.

  Motifs you are required to layout with its id, extents in m[width, height, depth], individual objects in the motif: """<MOTIFS>"""
  You are also given the floorplan and top down view of each scene motif.

  Follow these steps:

  1. Review Input Information:
    - Study the provided room analysis
    - Note door location and swing path
    - Identify any architectural features or constraints
    - Consider the room's designated purpose

  2. Position motifs:
    - For each scene motif, specify:
      - Precise center point coordinates (x, z) of the AABB bounding box within the room
      - Optimal rotation angle in degrees (counter-clockwise relative to the Z-axis) (default is 0 facing south)
      - Consider wall alignment for scene motifs that traditionally work best against walls

  3. Optimize Placement:
    - Position each motif to:
      - Align appropriate scene motifs flush with walls when possible, use common sense to determine if the scene motif should be aligned with a wall
      - Distribute scene motifs evenly throughout the available space and avoid cramping multiple scene motifs in the same area of the room
      - Avoid placing scene motifs too close to the door
    - Aim to fill each corner of the room with a scene motif

  Wall Alignment Considerations:
    - Most furniture typically goes against walls (like beds, sofas, or cabinets) unless otherwise specified:
      - Position the initial coordinates near your intended wall for optimal snapping
      - Use wall_alignment: true to enable wall snapping
      - Specify wall_alignment_id to target a specific wall (walls are numbered 0 to N-1 clockwise from room vertices)
      - The object will snap to the specified wall and rotate to face into the room
    
  Output Format:
  ```json
  {
    "positions": [
      {
        "rationale": "concise explanation for placement and orientation",
        "id": "unique identifier for each furniture group",
        "position": [x, z],
        "rotation": angle_in_degrees,
        "wall_alignment": boolean (true if the scene motif should be aligned with a wall, false otherwise),
        "wall_alignment_id": integer (index of the target wall, 0-based, counting clockwise from room vertices. Can be ignored if you do not want to align with a specific wall),
        "ignore_collision": boolean (USE WITH CAUTION: true if the scene motif should not be checked for collision with other furniture (e.g. rug on its own only), false otherwise)
      }
      // Repeat for each scene motif
    ]
  }
  ```
\end{minted}

If the occupancy of the room is below $\occthresh$, we prompt the VLM to suggest potential objects to add to the room.
We specifically instruct it to avoid existing objects that are already in the room for better diversity.
\begin{minted}[breaklines, breakafter=d, fontsize=\scriptsize]{yaml}
large_furniture_extra: >-
  The following are the large furniture that is already placed in the room: <LARGE_FURNITURE>.
  Given a input room description "<ROOM_TYPE>", 
  You are required to add a few more (1-3) large furniture according to the room type to fill the empty space.
  Remember that large furniture can only be placed on the floor,
  do not generate any objects that is placed on the wall (e.g. painting, mirror) or objects on top of the large furniture (e.g. a lamp on a table).
  Do not use the same furniture as the one already placed in the room (e.g. a desk in the room, do not generate another desk).

  For each large furniture that is required to be placed on the floor (e.g. a sofa, a bed, a cabinet, a desk, etc.) 
    1. Integer id of the piece
    2. The name of the piece (be speicific and sperate different categories of furniture, e.g. dining table, dining chair, office chair, etc.)
    3. The appearance/style description of the piece (be specific, e.g. "large wooden desk", "large round dining table")
    4. The dimensions of the piece [width, height, depth] in meters according to the description, give the most likely dimensions in meters
    5. The amount of the piece

  Critical requirements:
    - Each entry must represent a SINGLE type of object

  Respond with JSON:
  ```json
  {
    "objects": [
      {
        "id": large_furniture_id (integer),
        "name": "furniture_name",
        "description": "appearance/style description of the furniture",
        "dimensions": [width, height, depth],
        "amount": number of same furniture (integer),
      },
    ],
  }
  ```
\end{minted}

\subsection{Small Object Level Prompts}
\label{sec:supp/prompts/small_object_level}

We provide the small object level prompts below.
For small object placement, the VLM is further prompted to select which of the floor and wall-mounted objects should be populated with commonly co-occurring items.
Other procedure is similar to the one in the previous section.

\begin{minted}[breaklines, breakafter=d, fontsize=\scriptsize]{yaml}
system: >-
  You are an AI assistant specializing in realistic and functional small object placement on furniture surfaces. 
  Your task is to populate the surface of a furniture with small objects that are both functional and aesthetically pleasing.

  You are provided with a top-down 2D plot visualization that contains:
  - A highlighted area in the center representing the furniture surface to be populated
  - "X" markers indicating surrounding objects with their names
  - A red arrow indicating the front direction of the furniture
  - Axis measurements in meters showing the scale and dimensions
  - The exact dimensions of the small object labeled in the plot

  All 2D visualization uses:
  - The center of all plots is the origin (0,0)
  - X-axis: represents width (negative is left, positive is right)
  - Z-axis: represents depth (positive is back, negative is front)
  - Highlighted area: shows the usable surface for object placement
  - "X" markers: represent objects around the furniture
  - Black arrow: indicates the front-facing direction
  - Grid lines: help with precise measurements and positioning

  Follow these steps:

  1. Context Analysis:
    - Count and map ALL surrounding objects that indicate usage (e.g., chairs, benches)
    - Create a corresponding number of place settings or interaction points
    - Ensure EVERY surrounding object that needs interaction has a corresponding setup
    - Map potential interaction zones based on ALL surrounding object positions

  2. Surface Analysis:
    - Map the entire usable surface area in detail
    - Divide surface into equal sections based on number of surrounding objects
    - Ensure each interaction zone has adequate space
    - Reserve central area for shared items

  3. Object Selection & Distribution:
    - Place one complete set of required items for EACH identified interaction point
    - Distribute objects to ensure equal access from all interaction points
    - Ensure no interaction points are missed or doubled
    - Add shared items only after all required individual setups are complete

  4. Functional Optimization:
    - Ensure frequently used items remain accessible
    - Create clear paths for reaching objects
    - Account for object removal/replacement

  5. Rotation Specification:
    - For standalone objects: specify an "angle" in degrees (counterclockwise from positive Y-axis)
    - For objects that should face users: use "facing" with the ID of the relevant object
    - Examples of facing objects:
      * Place settings facing chair positions
      * Reading materials oriented toward seating
      * Control devices pointing toward user positions
      * Display items angled for optimal viewing
\end{minted}

\begin{minted}[breaklines, breakafter=d, fontsize=\scriptsize]{yaml}
populate_surface_motifs: >-
  Based on the object layered description earlier, 
  suggest object groupings only for the following small objects """<SMALL_OBJECTS>""" 
  based on the room description """<ROOM_TYPE>""" and the previous assignment of surfaces to objects.

  Critically assess if objects must be grouped. Group them only if they form a strong, not separable functional unit 
  (e.g., a computer next to a mouse) or have a direct, necessary spatial dependenancy (e.g., a cup on top of a saucer). 
  Otherwise, they should be in separate, smaller motifs or as single-object motifs. 
  Avoid grouping loosely related items even if they are nearby.
  
  For each motif, define:
    1. One or multiple objects (prefer single objects unless truly functionally dependent)
    2. Realistic dimensions and clear spatial relationships between pieces
    3. Total footprint dimensions [width, height, depth]
    4. Required clearance space in meters

  Guidelines for motif:
    - PREFER single-object motifs over multi-object groupings
    - ONLY group objects with explicit functional dependency (not thematic or decorative similarity)
    - Each small object can only be used once in each motif
    - Each motif can only contain a MAXIMUM of 4 different types of objects, if there are more than 4, split them into multiple motifs
    - If unsure whether to group, ALWAYS create separate motifs
    - Ensure ALL objects from the provided list are used across your arrangements
    - IMPORTANT: Do NOT create duplicate arrangements with identical compositions.
    - For multiple instances of the same motif type, use numbered suffixes (e.g., "plant_display_1", "plant_display_2", "book_stack_1", "book_stack_2")
    
  Respond with JSON with the following format:
  ```json
  {
    "arrangements": [
      {
        "rationale": "concise explanation of arrangement functionality",
        "id": "unique arrangement identifier (e.g. table_setup)",
        "area_name": "name of motif",
        "composition": {
          "description": "direct and precise description of local object relationships only without any style details, accurately reflecting the exact quantities specified (e.g. a stack of four books, a lamp, a fork and a knife on each side of a plate)",
          "furniture": [
            {"id": id1, "amount": integer}, (according to the description)
            ... (and more only if they are functionally related)
          ],
          "total_footprint": [width, height, depth],
          "clearance": clearance_in_meters,
        },
      },
      ... 
    ],
  }
  ```

  Description guidelines:
  - Do not include references to room features (walls, doors, windows) and objects that are not small objects in description.
  - Do not reference the furniture that the small objects are on.
  - Do not include non-spatial or stylistic relationships (e.g. design style details); only give concrete, spatial relationships.
  - The description must accurately reflect the exact quantities specified in the "amount" field
\end{minted}

\begin{minted}[breaklines, breakafter=d, fontsize=\scriptsize]{yaml}
describe_layered_object: >-
  Please describe in a few sentences on the geometry of the layered object """<LARGE_OBJECT>""" in a """<ROOM_TYPE>""" based on the following inputs:
  Think step by step about the geometry of the object and the space it occupies.
  - List of objects in the scene with object id, name, position and dimension: """<EXISTING_OBJECTS>"""
  - Object to be populated: """<OBJECT_TO_POPULATE>"""

  Image (Layer breakdown):
  - Shows """<LARGE_OBJECT>"""'s layers from highest (Layer 0) to lowest
  - Use this to understand the layer structure and available space
  - The position of each small object is relative to this image per layer
  - Each layer shows:
    - Height (y value in meters)
    - Available space above
    - Highlighted surfaces (available space)
\end{minted}

\begin{minted}[breaklines, breakafter=d, fontsize=\scriptsize]{yaml}
populate_object_layered: >-
  Given a scene motif """<PARENT_MOTIF_DESCRIPTION>""" in a room described as """<ROOM_TYPE>""",
  populate the specified object with appropriate small objects from the provided motifs.

  Small object motifs to place: """<SMALL_MOTIFS_TO_POPULATE>"""

  Layer information for small object motifs to place:
  - Each layer includes:
    - Layer index (starting from 0, highest surface first)
    - Layer height (from ground in meters)
    - Whether it's the topmost visible surface
    - Vertical space above the layer
    - Surface details:
      - Surface ID and color
      - Dimensions (width, depth)
      - Area
      - Center position [x, z]

  Layer Structure:
  """<LAYER_INFO>"""

  You are provided with two reference images:

  Image 1 (Top-down motif view):
  - Shows all large objects in context with front directions (black arrows)
  - Use this to reason about spatial relationships

  Image 2 (2D Layer breakdown for """<LARGE_OBJECT>"""):
  - Visualizes the surfaces of each layer, from top-left (highest) to bottom-right (lowest)
  - Small object positions are relative to this image
  - Each layer displays:
    - Height (y in meters)
    - Space above
    - Surface dimensions and availability

  Critical Requirements:
  1. You MUST populate ALL surfaces in ALL layers (even if empty, include empty arrays [])
  2. Use only the exact surface IDs provided in the layer information
  3. For rotation "facing" field, only reference objects within the same motif: <PARENT_MOTIF_OBJECTS>
  4. Each small object must reference a valid motif ID from the available motifs

  Guidelines:
  - Use only the layers and surface IDs provided
  - Place objects on surfaces likely to be used (avoid placing on very tall furniture tops unless typical)
  - Use available vertical space wisely; consider real-world usability

  Return the result in JSON:
  ```json
  {
    "large_object_name": {
      "layer_0": {
        "surface_0": [
          {
            "id": "motif_id_from_available_motifs",
            "position": [x, z],
            "rotation": { (choose one of the following)
              "angle": angle,  // Default: 0 degrees (facing the front direction of the parent object)
              "facing": "object_name_from_same_motif" // Use when object should face towards a specific object
              "face_away": "object_name_from_same_motif" // Use when object should face away from a specific object
            },
            "rationale": "Brief explanation"
          }
        ],
        "surface_1": []  // Empty if no objects, but must be included if there is surface_1 in the layer
      },
      other layers if any
    }
  }
  ```

  Placement Guidelines:
  - Consider real-world usage patterns
  - Respect vertical space constraints
  - Distribute objects logically across available surfaces
  - Leave some surfaces empty if appropriate, but include them as empty arrays

  Example:
  If layer_info shows dining table has layer_0 with surface_0 and surface_1:
  ```json
  {
    "dining table": {
      "layer_0": {
        "surface_0": [
          {
            "id": "place_setting_1",
            "position": [0.3, 0.3],
            "rotation": {"facing": "dining chair"},
            "rationale": "Positioned for the chair"
          }
        ],
        "surface_1": [
          {
            "id": "place_setting_2", 
            "position": [-0.3, -0.3],
            "rotation": {"angle": 0},
            "rationale": "Opposite side placement"
          }
        ]
      }
    }
  }
  ```
\end{minted}

\begin{minted}[breaklines, breakafter=d, fontsize=\scriptsize]{yaml}
small_objects_layered: >-
  You are required to populate only and exactly the following small objects """<SMALL_OBJECTS>"""
  on the following motif: """<MOTIF_DESCRIPTION>""",
  only on the following large furniture: """<LARGE_FURNITURE>""",
  in a room with description: """<ROOM_TYPE>""",

  The layer information is: """<LAYER_INFO>"""

  CRITICAL OBJECTIVE: You must place the EXACT total quantity specified for each object type across ALL furniture surfaces. This is a strict requirement - no more, no less.

  Quantity Distribution Strategy:
  1. First, identify all available surfaces across all layers
  2. Calculate how to distribute each object type to reach the exact total
  3. Ensure the sum of all amounts for each object type equals the required total exactly

  For each small object entry, provide:
    1. The name of the piece (use EXACTLY the same name as provided in the input)
    2. The appearance/style description of the piece (be specific, e.g. "glass cup")
    3. The dimensions of the piece [width, height, depth] in meters according to the description
    4. The amount of the piece (integer)

  JSON Structure Requirements:
    - Each large_object_name appears exactly once as a top-level key
    - Each layer_X appears exactly once under each large_object_name
    - Each surface_X appears exactly once under each layer_X
    - All layers and surfaces from the layer information must be included
    - CRITICAL: Do NOT duplicate layer keys (e.g., do not define "layer_0" multiple times)

  Additional Requirements:
    - Each entry must represent a SINGLE type of object, do not generate composite sets
    - If there are multiple small objects with different appearance/size, break them into multiple entries
    - Use exact number of layers and surface IDs from layer information
    - Consider space available on each surface and height of each layer (layer_0 is highest)

  Respond with JSON:
  ```json
  {
    "large_object_name": {
      "layer_0": {
        "surface_0": [
          {
            "name": "small_object_name",
            "description": "appearance/style description of a single small object (e.g. glass cup)",
            "dimensions": [width, height, depth],
            "amount": number of same small object (integer),
          },
        ]
      },
      "layer_1": {
        "surface_0": [
          {
            "name": "small_object_name",
            "description": "appearance/style description of a single small object",
            "dimensions": [width, height, depth],
            "amount": number of same small object (integer),
          },
        ]
      },
      ...
    },
    ...
  }
  ```
\end{minted}

\begin{minted}[breaklines, breakafter=d, fontsize=\scriptsize]{yaml}
choose_objects: >-
  Choose from the following list of objects: """<OBJECT_LIST>""" 
  Which objects usually has small objects placed on top/inside of it?
  
  You can respond with an empty array if it is absolutely certain that all objects are not meant to have any objects placed on top of them.
  Respond exactly with the object names in JSON format.
  ```json
  {"objects": ["object_1", "object_n", ...]} 
  ```
\end{minted}

\end{document}